\documentclass[aps,pre,twocolumn,superscriptaddress]{revtex4-2}
\usepackage{amssymb}
\usepackage{amsmath,bm}
\usepackage{graphicx,color}
\usepackage{bbm}
\usepackage[bottom]{footmisc}
\usepackage{multirow}
\usepackage{xcolor}
\usepackage{xpatch} %
\makeatletter   
\usepackage{xpatch} %
\usepackage{cancel}
\usepackage{appendix}

\newcommand{\B}[1]{{\bm{#1}}}

	\newcommand{\Lag}{\mathcal{L}}
	
	\newcommand{\dif}{\mathrm{d}}

	\newcommand{\C}[1]{{\mathcal{#1}}}
	
	\DeclareMathOperator{\sech}{sech}
	
\begin{document}
		\title{Analytic Nonlinear Theory of Shear Banding in Amorphous Solids}
		\author{Avanish Kumar}
		\affiliation{Sino-Europe Complex Science Center, School of Mathematics, North University of China, Shanxi, Taiyuan 030051, China}
		\author{Itamar Procaccia} 
		\affiliation{Hangzhou International Innovation Institute, Beihang University, Hangzhou, 311115 China}
		\affiliation{Sino-Europe Complex Science Center, School of Mathematics, North University of China, Shanxi, Taiyuan 030051, China}
		\affiliation{Dept. of Chemical Physics, The Weizmann Institute of Science, Rehovot 76100, Israel}
		
		\begin{abstract}
			The aim of this paper is to offer an analytic theory of the shear banding instability in amorphous solids that are subjected to athermal quasi-static shear. To this aim we derive nonlinear equations for the displacement field, including the consequences of plastic deformation on the mechanical response of amorphous solids. The plastic events collectively induce distributed dipoles that are responsible for screening effects and the creation of typical length-scales that are absent in classical elasticity theory. The nonlinear theory exposes an instability that results in the creation of shear bands.  By solving the weakly nonlinear amplitude equation we present analytic expressions for the displacement fields that is associated with shear bands, explaining the role of the elastic moduli that determine the width of a shear band from ductile to brittle characteristics. We derive an energy functional whose Hessian possesses an eigenvalue that goes to zero at the shear-banding instability, providing a prediction for the critical value of the accumulated stress that results in an instability. 
		\end{abstract}
		\maketitle
		
		\section{Introduction}
		
		Understanding the mechanical response of amorphous solids, like glasses and granular media, require considerations that go beyond standard approaches to elasticity theory and solid mechanics. To start,  contrary to perfect crystalline solids that can exhibit purely elastic responses to strain, amorphous solids suffer from plastic responses for any amount of strain, and these typically appear as quadrupolar ``Eshelby inclusions" \cite{54Esh,99ML,06ML}. Physically, this means that constituent particles can move around in so called  ``non-affine" processes, not respecting the picture of affine changes of coordinates that are at the basis continuum mechanics which employs curvilinear coordinate systems, co- and contravariant tensors, Christoffel symbols, Mohr’s circles, etc. \cite{82Gur,00HH}. Amorphous solids lack long-range order, and they do not possess a unique ``ground state". They can be cooled down to zero temperature, where they can reside in one of many available local equilibria, which under mechanical strains can easily exchange relative stability. Research in the last decade or two indicated that classical elasticity theory needs to be reconsidered for the treatment of amorphous solids.
		
		In our approach we are interested in the response of an amorphous solid that is initially in mechanical equilibrium at zero temperature (i.e having zero resultant force on each particle), which is then subjected to strain, and then relaxed back to mechanical equilibrium at zero temperature by energy minimization. Being aware that this is far from covering the most general mechanical response which can be time-dependent or effected by temperature fluctuations, we discover that this modest protocol exposes a myriad  interesting new phenomena that are worthy of detailed study before opening up to effects of strain rate and temperature. We also realize that the naturally measured field for our purposes, whenever possible, is the displacement field (both in numerics and in experiments). We extract the displacement field from the positions of the particles before and after the strain and energy minimization.
		We define the system strain and stress tensors from the displacement field, and thus never run into issues of compatibility that can come up in other formulations of the problem. 
		
		A major aim of this paper is to explain and provide a theory for the phenomenon of shear banding. Shear banding is a generic and common instability prevalent in amorphous solids from metallic glasses to computer simulations of granular matter \cite{82SPH,07SKLF,26FJP}. The phenomenon was studied extensively in numerical simulations using quasi-static protocols, in which the external load, like shear strain, is added in small steps after each of which the system is brought back to mechanical equilibrium by energy gradient minimization or damped dynamics \cite{20BLLVP,18OBBRT}. Brittle materials exhibit a sharp instability at a critical value of the accumulated stress. Here the displacement field shows a sharp jump across a line in 2-dimensions and across a plane in 3-dimension, concentrating the shear along this separating line or plane. In more ductile materials the shear is less localized, showing a wider and smoother shear band, but nevertheless still clearly noticeable. Our aim here is to offer a theory to predict the critical accumulated stress that results in instability,  provide an analytic solution of the displacement field profile during the shear band formation, and understand the material properties that distinguish brittle from ductile materials. 
		
		In a series of recent publications, it was discovered that the prevalence of plastic events in the mechanical response to loads in amorphous solids results in screening phenomena that are akin, but richer and different, to screening effects in electrostatics \cite{21LMMPRS,22MMPRSZ,22BMP,22KMPS}. Plastic events, which are localized quadrupole singularities in a reference curvature field \cite{15KMS, 15MLAKS},
		and consequently typically quadrupoles in the displacement field, can act as screening charges. It was shown that when the density of plastic quadrupoles is uniform, their effect is limited to renormalizing the elastic moduli, but the structure of linear elasticity theory remains intact. But when the quadrupoles density
		becomes non-uniform,  effective dipoles defined by the gradients of the quadrupole density, cannot be neglected. The presence of effective dipoles has surprising
		consequences, changing the analytic form of the response
		to strains in ways that cannot possibly be predicted by
		standard elasticity theory. It was concluded that one needs to consider a new theory, and this emergent theory was confirmed by comparing its predictions to results of extensive experiments and simulations
		\cite{21LMMPRS,22MMPRSZ,22BMP,22KMPS}.
		
		This progress cannot be applied as is to the shear banding phenomenon, since it was based on linear approximations. To make the issue clear, all the above cited works are based on the linear relation between strain field $\B u$ and displacement field $\B d$, i.e.
		\begin{equation}
			u_{\alpha\beta} = \frac{1}{2}\left(\partial_\alpha  d_\beta  + \partial_\beta d_\alpha \right) \ .
			\label{defu}
		\end{equation} 
		In addition, all the Lagrangians used in previous work were expanded to quadratic order in the relevant fields, providing new and interesting equations for the displacement field $\B d$, but keeping them to linear order. This will not be sufficient for our present aims. We will argue below that there are two types of non-linearity that need to be taken into account, each having a different role. The first is in the relation between strain and displacement, i.e.
		\begin{equation}
			u_{\alpha\beta}= \frac{1}{2}\left[\partial_\alpha  d_\beta + \partial_\beta d_\alpha  \right] + \frac{1}{2} \partial_\beta d_\gamma\partial_\alpha d_\gamma \ .
			\label{defunl}
		\end{equation}
		Secondly, we will need to expand our Lagrangians to higher orders in the relevant fields, as we explain in detail in the present paper. These two extensions of the linear theory are essential, and are in clear distinction with previous approaches to the shear banding problem, as done for example in Refs. \cite{75RR,80RR}. In these earlier works no attention was paid to the topological dipole fields or to the nonlinear stress-strain relationship. Therefore previous approaches could not come up with a solution of the shear-band profile. We are in position to do this since
		we will derive
		{\em nonlinear} equations for the displacement field, and will face the task of solving these nonlinear equation with the appropriate boundary conditions. We will see that this program will end up with a proper theory of shear banding.
		
		The structure of this paper is as follows: Sect.~\ref{linear} will provide a summary of the linear theory. The consequences of Eq.~(\ref{defunl}) will be explored in Sect.~\ref{unonlin}.  Section \ref{nonlinP} is devoted to the derivation of the nonlinear equation for the displacement field, ending up with Eq.~(\ref{finalNL}). This equation is the basis for the rest of the paper. The analysis of this equation is offered in Sect.~\ref{anal}, in which we present the strategy of finding the solution of the shear-band problem. We derive an effective equation for this solution which is analyzed in Sect.~\ref{effective}. Here we propose an energy functional whose Hessian has a lowest eigenvalue that vanishes at the shear-banding instability. The onset of the instability where the critical value of the background stress is identified is discussed in Sect.~\ref{onset}.
		The solution of the nonlinear equation for the displacement field is provided in Sect.~\ref{tanh}. Shear localization and its connection to the nonlinear Hessian is discussed in Sect.~\ref{sloc}. Finally, the new results are summarized and discussed further in Sect.~\ref{summary}. In this concluding section we also compare and contrast our theory with previously available theories of shear banding.  
		
		\section{The Linear Theory}
		\label{linear}
		
		In this section we provide a summary of the theory of the mechanical response of amorphous solids to external loads that was developed until now. It is impossible to provide full details, but the material is extensively published and the novice reader is advised to consult the provided references. We will develop here the theory for two spatial dimensions, but the extension to 3-dimensions is available, cf. Ref.~\cite{23CMP}. 
		
		Classical elasticity can be derived from a Lagrangian $	U_\text{el}$ which takes into account the strain and stress fields. Defining the strain field Eq.~(\ref{defu})
		and the stress field $\sigma_{\alpha\beta}= A^{\alpha\beta\gamma\delta} u_{\gamma\delta}$, we write
		\begin{equation}
			U_\text{el} =\!\!\int \dif^2 x \frac{1}{2} A^{\alpha\beta\gamma\delta} u_{\alpha\beta}u_{\gamma\delta} \ .
		\end{equation}
		Here $A^{\alpha\beta\gamma\delta}$ is the standard tensor of elastic moduli, whose explicit form in homogeneous isotropic systems is given in Eq.~(\ref{elten}) below. Upon minimizing this Lagrangian one derives the equilibrium equation of classical elasticity
		\begin{equation}
			\partial_\alpha  \sigma^{\alpha\beta} = 0 \ ,
			\label{Eq}
		\end{equation}
		which is solved subject to the boundary conditions. Using the linear relations between stress, strain and displacement, this equation translates to an eqivalent equation for the displacement field which is available in
		all texts of classical elasticity \cite{Landau},
		\begin{equation}
			\Delta \mathbf{d} + \left(1+\frac{\lambda}{\mu}\right) \nabla \left(\nabla\cdot \mathbf{d}\right) = 0 \ .
			\label{elastic}
		\end{equation} 
		
		Next, it was realized that in amorphous solids the application of external loads always results, in the thermodynamic limit, in plastic deformations \cite{10KLP,11HKLP}. These are generically appearing as quadrupoles in the displacement field, sometime referred to as ``Eshelby inclusions" \cite{54Esh,99ML,06ML}.  To derive equations for the displacement field $\B d (\B r)$ in the presence of quadrupoles, we start with an expression for the Lagrangian of the system, consistent with the underlying symmetries, presented up to quadratic order in the relevant fields. 
		The mechanical energy stored in the system stems from three main contributions \cite{13DHP}. First, the energetic cost associated with the bare imposed stress field $U_\text{el}$. Second is the interaction of the induced quadrupoles with the elastic background $U_\text{Q-el}$. Lastly, there is the self-interaction of the quadrupoles, reflecting their nucleation cost. Explicitly, 
		$U = U_\text{el} + U_\text{Q-el} + U_\text{QQ}$.
		\begin{eqnarray}
			&&	U_\text{Q-el} =  \int \dif^2 x \Gamma^{\alpha\beta}_{\gamma\delta} u_{\alpha\beta}Q^{\gamma\delta}\ , \nonumber \\
			&&U_\text{QQ} =\int \dif^2 x \mathcal{F} \left(Q^{\alpha\beta},\partial_\beta Q^{\alpha\beta}\cdots\right)\ . 
			\label{eq:energydecomp}
		\end{eqnarray}
		Here $\B \Gamma$ is an appropriate coupling tensor, that eventually renormalizes the standard moduli.  $\mathcal{F}$ represents the energy cost of the induced plastic quadrupoles, including their first and higher gradient terms. When the gradient terms are important they lead to screening, and see below for details. 
		\subsection{The quasi-elastic regime}
		In the quasi-uniform quadrupolar field limit, corresponding to large
		energetic cost for nucleating dipoles, quadrupoles vary
		slowly in space to avoid effective dipoles, hence \cite{21LMMPRS,23CMP}
		$\C F=\mathcal{F} \left(Q^{\alpha\beta}\right) = \frac{1}{2}\Lambda_{\alpha\beta\gamma\delta} Q^{\alpha\beta}Q^{\gamma\delta}	$. Upon minimizing $U$ with respect to the fundamental fields $d$ and $Q$, using  \eqref{eq:energydecomp}, we find  \cite{21LMMPRS,23CMP}
		a linear screening relation (analogous to the linear relation between electric field and induced polarization in dielectric materials \cite{49Fro})
		\begin{equation}
			Q^{\alpha\beta} = - \Lambda^{\alpha\beta\mu\nu}  \Gamma_{\mu\nu}^{\gamma\delta} u_{\gamma\delta} \equiv  - \tilde{\Lambda}^{\alpha\beta\gamma\delta} u_{\gamma\delta} \ ,
			\label{eq:Screeningrelation}
		\end{equation}
		where $ \Lambda^{\alpha\beta\mu\nu}$ is the inverse of $\Lambda_{\alpha\beta\mu\nu}$. 
		The second result that one finds is
		\begin{equation}
			\partial_\alpha  \tilde{\sigma}^{\alpha\beta} = 0 \ ,
			\label{eq:Equilibrium}
		\end{equation}
		where $\tilde \sigma^{\alpha\beta} \equiv \sigma^{\alpha\beta}+ \Gamma^{\alpha\beta}_{\gamma\delta} Q^{\gamma\delta}$.
		We see that the re-normalization of the quadrupole-quadrupole interactions results in a linear constitutive relation between inducing stress and induced quadrupoles which then renormalizes the elastic tensor \cite{20NWRZBC}. This is the analog of the situation in dielectrics, where the dielectric constant is renormalized
		by the induced dipoles \cite{49Fro}. Explicitly, the tensor of moduli is renormalized as follows:
		\begin{equation}
			\tilde A^{\mu\nu\rho\sigma}\equiv A^{\mu\nu\rho\sigma}+\Lambda_{\alpha\beta\gamma\delta}\tilde{\Lambda}^{\alpha\beta\mu\nu}\tilde{\Lambda}^{\gamma\delta\rho\sigma}-2\Gamma_{\mu\nu}^{\gamma\delta}\tilde{\Lambda}^{\gamma\delta\rho\sigma} \ .
			\label{Arenor}
		\end{equation}
		Using this renormalized tensor, the Lagrangian in the quasi-elastic regime
		can be written again in the form 
		\begin{equation}
			{\cal L}\equiv	\frac{1}{2}\tilde A^{\alpha\beta\gamma\delta} u_{\alpha\beta}u_{\gamma\delta} \ ,
			\label{renL}
		\end{equation}
		leaving the form of the theory unchanged. 
		\subsection{The screening regime}
		\label{highdensity}
		When the quadrupole field in not uniform, one cannot neglect the gradient terms. Using the renormalized elastic tensor Eq.~(\ref{Arenor}), we now consider the gradient terms in the function $\mathcal{F} \left(Q^{\alpha\beta},\partial_\beta Q^{\alpha\beta}\cdots\right)$, $U=\int d^2x \Lag$ and the Lagrangian $\Lag$ reads \cite{21LMMPRS,23CMP}:
		\begin{equation}
			\begin{split}
				\Lag &=  \frac{1}{2} \tilde{A}^{\mu\nu\rho\sigma} u_{\mu\nu}u_{\rho\sigma} + 
				\frac{1}{2} \Lambda_{\alpha\beta} \partial_\mu Q^{\mu\alpha}  \partial_\nu Q^{\nu\beta}
				+ \Gamma_{\alpha}^{\,\,\beta} \partial_\mu Q^{\mu\alpha} d_{\beta} \ .
			\end{split}
			\label{lagdip}
		\end{equation} 
		the last term is the only additional quadratic term allowed by symmetry. Note that the quadrupole-quadrupole terms were not included since the renormalization of the moduli is already taken into account.

		Denoting the gradients on the quadrupoles as effectively induced dipoles  $P^\alpha \equiv \partial_\beta Q^{\alpha\beta}$, and minimizing with respect to the fundamental fields $d$ and $Q$ we find 
		\begin{eqnarray}
			\partial_\alpha[	P^{\alpha} +  \Lambda^{\alpha\beta}  \Gamma_{\beta}^{\gamma} d_{\gamma}]&=& 0  ,\\ 
			\partial_\alpha \sigma^{\alpha\beta} &=& \Gamma_{\alpha}^{\beta} P^{\alpha} = -\Gamma_{\alpha}^{\beta}  \Lambda^{\alpha\mu}  \Gamma_{\mu}^{\gamma} d_{\gamma} 
			\ .
			\label{ScreeningEq2}
		\end{eqnarray}
		We note that Eq. (12) means that $P^{\alpha} +  \Lambda^{\alpha\beta}  \Gamma_{\beta}^{\gamma} d_{\gamma}=C$, and by choosing $C=0$ we fix the gauge and find as a result a relationship between the dipole field and the displacement field,
		\begin{equation}
			P^{\alpha} =- \Lambda^{\alpha\beta}  \Gamma_{\beta}^{\gamma} d_{\gamma}	 \ .
			\label{Pvsd}
		\end{equation}
		At this point we use again the linear relations between stress, strain and displacement, to rewrite Eq.(\ref{ScreeningEq2}) in terms of the displacement field. The final equation reads
	\begin{equation}\label{L0}
		\bf  \Delta  {d} + (\tilde{\lambda} +1)\nabla (\nabla \cdot  {d}) + \widetilde K {d} =0,
	\end{equation}
	where $\B {\widetilde K}$ has the following form \cite{24FHKKP,25KPS,25CSWDM},
	\begin{align}\label{L1}
		\B {\widetilde K} \equiv \B \Gamma \B \Lambda \B \Gamma =
		\begin{bmatrix}
			\bf    \tilde{\kappa}_{e}^{2} & \bf-\tilde{\kappa}_{o}^{2}  \\
			\bf    \tilde{\kappa}_{o}^{2} &\bf \tilde{\kappa}_{e}^{2} 
		\end{bmatrix} = \frac{\B K}{\mu},
	\end{align}
	where, $\B K$ is the screening matrix (SM) with the following screening parameters, 
	\begin{align}\label{SM}
		\B { K}  =
		\begin{bmatrix}
			\bf     \kappa_{e}^{2} & \bf-\kappa_{o}^{2}  \\
			\bf    \kappa_{o}^{2} &\bf \kappa_{e}^{2} 
		\end{bmatrix}~~ \text{such that} ~~ \bf \kappa_{e}^{2} = \mu ~ \tilde{\kappa}_{e}^{2},~ \kappa_{o}^{2} = \mu ~ \tilde{\kappa}_{o}^{2} . \tag{SM}
	\end{align}
		The screening parameters, both the diagonal (which lead to translational symmetry breaking) and off-diagonal (which lead to chiral symmetry breaking), are a-priori predictable to high precision from the properties of the solutions of the governing Eq.~(\ref{L0}). 
		Ideas of how to predict a-priori the numerical value of the emergent inverse scales can be found in \cite{24JPS,25PS}. 
		
		\section{Nonlinear Strain-displacement relation}
		\label{unonlin}
		
		To proceed with the nonlinear theory, we need to worry about three aspects. The first is the consequence of the nonlinear relation between strain and displacement, Eq.~(\ref{defunl}). The second is the expansion of the Lagrangian Eq.~(\ref{ScreeningEq2}) to higher orders in the dipole field $P^\alpha \equiv \partial_\beta Q^{\alpha\beta}$. This is done in the next section.  The third is the consequence of having a background stress field that is non-zero. Let us first deal with the last one.
		
		As explained in the introduction, we explore here not a single step of load, but a quasi-static protocol that creates a substantial background stress $\Sigma^{\alpha\beta}$ which was not taken into account previously. 
		In order to do it, here we will use the fact that in quasi-static protocols, prior to the studied stress drop (which may have ignited a shear band) the system underwent energy minimization, reached a mechanical equilibrium state with $\partial_\alpha \Sigma^{\alpha\beta}=0$. Since the next step of strain  is small, we can justify the linear approximations $\sigma^{\alpha\beta} =\tilde A^{\alpha\beta\gamma\delta} u_{\gamma\delta}$ for the associated stress, where $\tilde A^{\alpha\beta\gamma\delta}$ is the renormalized elastic tensor that is relevant at the present state of the system. 
		Note that here $\Sigma^{\alpha \beta}$ is the accumulated background stress, and $\sigma^{\alpha \beta}$ is the change of stress during the plastic event.
		
		Let us then discuss the first aspect, with the energy including the background 
		stress field and the quadrupolar contributions,
		\begin{eqnarray}
			U& =& \frac{1}{2} \int d^2x ~(\Sigma^{\alpha\beta}u_{\alpha\beta} + \sigma^{\alpha\beta}u_{\alpha\beta}) \\ &+&\frac{1}{2} \int d^2x ~\Lambda_{\alpha\beta\gamma\delta}Q^{\alpha\beta}Q^{\gamma\delta}+\int d^2x ~\Gamma_{\gamma\delta}^{\alpha\beta}u_{\alpha\beta}~Q^{\gamma\delta} \ .\nonumber
			\label{LQ5}
		\end{eqnarray}
		
		We minimize the energy by varying with respect to the displacement field $\B d$ and the quadrupolar field $\B Q$. For the variation with respect to $\B d$, using Eq.~(\ref{defunl}), we write
		\begin{align}\label{NLE21}
			\delta u_{\alpha\beta} = \frac{1}{2}\left[\delta d_{\alpha,\beta} + \delta d_{\beta,\alpha}  \right] + \frac{1}{2} \left[ d_{\gamma,\alpha} \delta  d_{\gamma,\beta} +  \delta d_{\gamma,\alpha} d_{\gamma, \beta}\right] ,
		\end{align}
		where we use here and below the notation $d_{\gamma,\alpha} \equiv \partial_\alpha d_\gamma$. In Appendix \ref{derive1} we display the full derivation leading to the equation
		\begin{align}\label{LQ13}
			\Sigma^{\alpha\beta}( d_{k,\beta})_{, \alpha} + \tilde{\sigma}_{\alpha k, \alpha}   +   (\tilde{\sigma}_{\alpha \beta}d_{k,\beta})_{, \alpha} =0. 
		\end{align}
		We see that this equation differs from its classical counterpart Eq.~(\ref{Eq}) by three aspects: (i) we gained a term proportional to the accumulated stress and (ii) another term
		due to the nonlinear relation Eq.~(\ref{defunl}).  Lastly (iii) the stress field $\tilde{\B \sigma}$ is normalized by the effects of the quadrupolar field, and cf. the explicit definition in Eq.~(\ref{Arenor}).  
		
		\section{Nonlinear Dipole contribution}
		\label{nonlinP}
		So far our Lagrangians were expanded up to quadratic order in the displacement,  quadrupolar and dipolar fields, yielding the linear equation in the displacement field Eq.~(\ref{L0}). At this point we are ready to expand further, and face the question which nonlinearities should be added. We have learned up to now that the quadrupolar fields by themselves can only renormalize the elastic moduli but do not change the structure of the theory. The dipolar fields are those that we need to focus on. Beyond the quadratic order term in the dipolar fields the next one that is allowed by symmetry is the quartic contribution, and we therefore write the Lagrangian 
		\begin{eqnarray}\label{QP1}
			\mathcal{L} &=& \frac{1}{2}(\Sigma^{\alpha\beta}u_{\alpha\beta} + \sigma^{\alpha\beta}u_{\alpha\beta}) + \Gamma_{\beta}^{\alpha}d_{\alpha}P^{\beta} + \frac{1}{2}\Lambda_{\alpha\beta}P^{\alpha}P^{\beta}\nonumber\\ &+&  \frac{1}{4} G_{\alpha \beta \gamma \delta} P^{\alpha}P^{\beta}P^{\gamma} P^{\delta} \ .
		\end{eqnarray}
		upon minimizing the energy $U\equiv \int d^2 x 	\mathcal{L} $ with respect to the fundamental fields $\B d$ and $\B Q$ we find
		\begin{align}\label{QP3}
			\partial_{Q}U =\int d^2x\left( \Gamma_{\beta}^{\alpha}  d_{\alpha} + \Lambda_{\alpha\beta}P^{\alpha} + G_{\alpha \beta \gamma \delta} P^{\alpha}P^{\gamma} P^{\delta} \right) \partial_\alpha\delta Q^{\alpha\beta}
		\end{align}
		\begin{align}\label{QP4}
			\partial_{d}U = \int d^2x (\Sigma^{\alpha\beta} +\sigma^{\alpha\beta})\delta u_{\alpha\beta}  + \Gamma_{\beta}^{\alpha}P^{\beta}\delta d_{\alpha}  .
		\end{align}
		From Eqn.~(\ref{QP3}) we obtain after gauge fixing
		\begin{align}\label{QP5}
			\Gamma_{\beta}^{\alpha}  d_{\alpha} + \Lambda_{\alpha\beta}P^{\alpha}  + G_{\alpha \beta \gamma \delta} P^{\alpha}P^{\gamma} P^{\delta} =0 .
		\end{align}
		This is a cubic equation in the dipole field $\B P$. We first solve the above equation for $\B P$ in the first-order approximation by neglecting the higher-order terms in $P$.
		\begin{align}\label{QP6}
			P^{\alpha} = -\Lambda^{\alpha\beta}\Gamma_{\beta}^{\eta} d_\eta \ .
		\end{align}
		
		From Eq.~(\ref{QP5}),
		\begin{align}\label{QP7}
			\Lambda_{\alpha\beta}P^{\alpha} = -\Gamma_{\beta}^{\delta}  d_{\delta}  - G_{\beta \gamma \delta \eta} P^{\gamma} P^{\delta}P^{\eta}, 
		\end{align}
		or
		\begin{align}\label{QP8}
			P^{\alpha} = -\Lambda^{\alpha\beta}\Gamma_{\beta}^{\delta}  d_{\delta} -  \Lambda^{\alpha\beta} G_{\beta \gamma \delta \eta} P^{\gamma} P^{\delta}P^{\eta} \ . 
		\end{align}
		Substituting the expressions for $\B P$ from Eq.~(\ref{QP6}) into Eq.~(\ref{QP8}), we obtain
		\begin{eqnarray}\label{QP9}
			P^{\alpha}& =& -\Lambda^{\alpha\beta}\Gamma_{\beta}^{\delta}  d_{\delta}\\ &-&\Lambda^{\alpha\beta} G_{\beta \gamma \delta \eta}  (-\Lambda^{\gamma \mu}\Gamma_{\mu}^{r} d_{r})  (-\Lambda^{\delta \nu}\Gamma_{\nu}^{p} d_{p})  (-\Lambda^{\eta \tau}\Gamma_{\tau}^{q} d_{q}) \ , \nonumber
		\end{eqnarray}
		which can be written as
		\begin{eqnarray}\label{QP10}
			P^{\alpha} &= &  -  \Lambda^{\alpha\beta}\Gamma_{\beta}^{\zeta}   d_{\zeta}\\  &+&  \Lambda^{\alpha\beta} G_{\beta \gamma \delta \eta} \Lambda^{\gamma \mu}\Gamma_{\mu}^{r} \Lambda^{\delta \nu}\Gamma_{\nu}^{p} \Lambda^{\eta \tau}\Gamma_{\tau}^q d_{p} d_{q} d_{r} \ .  \nonumber
		\end{eqnarray}
		
		Next we analyse Eq.~(\ref{QP4}), using Eq.~(\ref{defunl}) . 
		Performing the integration by parts we find
		\begin{widetext}
			\begin{equation}\label{QP13}
				\delta_d U =  -\int d^2 x  ~[\partial_\beta \sigma^{k\beta}   +   (\sigma^{\alpha\beta}d_{k,\alpha})_{, \beta}] ~ \delta d_{k}  +  \int d^2x~ \Gamma_{\beta}^{\alpha}P^{\beta}\delta d_{\alpha} 
				- \int d^2x~ [\cancelto{0}{\partial_\beta\Sigma^{k\beta}}   +   (\Sigma^{\alpha\beta}d_{k,\alpha})_{, \beta}] ~ \delta d_{k} +\text {Boundary terms}=0 \ . 
			\end{equation}
			
			Since the boundary terms must vanish, and the background stress is divergence-less, the above equation simplifies to the following equation,
			\begin{align}\label{QP15A}
				\Sigma^{\alpha\beta}(d_{k,\beta})_{, \alpha} + \sigma_{\alpha k, \alpha}   +   ({\sigma}_{\alpha \beta}d_{k,\beta})_{, \alpha} =  \Gamma_{\alpha}^{k} P^{\alpha} \ .
			\end{align}
			Substituting the expression of $P^{\alpha}$ from Eqn.~(\ref{QP10}) we obtain,
			
			\begin{align}\label{QP16}
				\Sigma^{\alpha\beta}(d_{k,\beta})_{, \alpha} + \sigma_{\alpha k, \alpha}   +   ({\sigma}_{\alpha \beta}d_{k,\beta})_{, \alpha} =  \Gamma_{\alpha}^{k} \left[ - \left\lbrace \Lambda^{\alpha\beta}\Gamma_{\beta}^{\zeta} \right\rbrace  d_{\zeta}  + \left\lbrace \Lambda^{\alpha\beta} G_{\beta \gamma \delta \eta} \Lambda^{\gamma \mu}\Gamma_{\mu}^{r} \Lambda^{\delta \nu}\Gamma_{\nu}^{p} \Lambda^{\eta \tau}\Gamma_{\tau}^{q}\right\rbrace d_{p} d_{q} d_{r} \right]\ . 
			\end{align}
			
			Finally, we rewrite the obtained equation in a more compact notation, 
			\begin{equation}\label{finalNL}
				\boxed{ \Sigma^{\alpha\beta}(d_{k,\beta})_{, \alpha} + \sigma_{\alpha k, \alpha}   +   ({\sigma}_{\alpha \beta}d_{k,\beta})_{, \alpha} = -L^{k\zeta} d_{\zeta} + T^{kpqr}d_p d_q d_r} \ ,	
			\end{equation}
		\end{widetext}
		where
		\begin{align}\label{QP26}
			L^{k\zeta} &= \Gamma_{\alpha}^{k}\Lambda^{\alpha\beta}\Gamma_{\beta}^{\zeta}, \\ \nonumber  
			T^{kpqr} &= \Gamma_{\alpha}^{k} \Lambda^{\alpha\beta} G_{\beta \gamma \delta \eta} \Lambda^{\gamma \mu}\Gamma_{\mu}^{r} \Lambda^{\delta \nu}\Gamma_{\nu}^{p} \Lambda^{\eta \tau}\Gamma_{\tau}^{q}.
		\end{align}
		Eq.~(\ref{finalNL}) is the nonlinear equation that will be analyzed in the rest of this paper.
		Note that if the connection between $\B L$ and the screening parameters continues to be valid in the nonlinear context, than $\B L$ can be estimated as $\B K$ of Eq.~(\ref{L1}).  
		
		The first step in the analysis would be to rewrite it in terms of the displacement field only. 
		\section{Analysis of the Nonlinear Equation Eq.~(\ref{finalNL}) }
		\label{anal}
		
		We are now faced with the task of analyzing Eq.~(\ref{finalNL}). Presently it has a mix of stress and displacement fields, so we will first rewrite it in terms of displacement field only. 
		\subsection{Extracting the displacement field}
		
		To this aim we rewrite Eq.~(\ref{defunl}) in the form
		\begin{equation}\label{QP29}
			\sigma_{\alpha\beta} =\widetilde A_{\alpha\beta\gamma\delta}\, \tfrac12\left( d_{\gamma,\delta} +d_{\delta,\gamma} +d_{\mu,\gamma}d_{\mu,\delta} \right).
		\end{equation}
		Substituting this in Eq.~(\ref{finalNL}), we find, after a somewhat lengthy but straightforward calculation, 
		\begin{widetext}
			\begin{eqnarray}\label{LHSk}
				&& \Sigma^{\alpha\beta}(d_{k,\beta})_{, \alpha} +\widetilde A_{\alpha k\gamma\delta}\,d_{\gamma,\delta\alpha} + \tfrac12 \widetilde A_{\alpha k\gamma\delta} \big(d_{\mu,\gamma\alpha}d_{\mu,\delta} +d_{\mu,\gamma}d_{\mu,\delta\alpha}\big)  
				+ \widetilde A_{\alpha\beta\gamma\delta}\Big( d_{\gamma,\delta\alpha}\,d_{k,\beta} +d_{\gamma,\delta}\,d_{k,\beta\alpha} \Big) 
				+ \tfrac12 \widetilde A_{\alpha\beta\gamma\delta}\big( d_{\mu,\gamma\alpha}d_{\mu,\delta}d_{k,\beta}\nonumber \\&& + d_{\mu,\gamma}d_{\mu,\delta\alpha}d_{k,\beta} + d_{\mu,\gamma}d_{\mu,\delta}d_{k,\beta\alpha} \big) = -L^{k\zeta} d_{\zeta} + T^{kpqr}d_p d_q d_r \ .
			\end{eqnarray} 
		\end{widetext}
		Evidently, this is a rather complex differential equation for the vector function $\B d$, requiring a numerical solution with the appropriate boundary conditions. Instead of doing this, we will proceed to explore the existence of a shear band solution, according to the following strategy.

		\subsection{Proposed strategy}
		The proposed strategy is composed of three steps: (i) we expect that the instability resulting in the formation of a shear band  will be identified with the existence of a soft mode of an appropriately defined Hessian, allowing us to project the displacement field onto that solution, making it effectively a scalar function of one variable. (ii) We will therefore rewrite Eq.~(\ref{LHSk}) for a scalar function of one variable. This is the meaning of the following ansatz Eq.~(\ref{defxi}). Since this ansatz is crucial for the theory below, we derive it from first principles in Appendix \ref{ansatz}.  (iii) Next we will find an energy functional whose Euler-Lagrange equations are the same as the said equation, allowing us to explicitly find the Hessian whose soft eigenfunction identifies the shear-band solution. 
		
		\subsection{Simplifying the equation for a scalar function of one variable}
		
		Let $\mathbf{x}=(x,y)$ denotes the spatial coordinates, and assume that the gradient of the displacement field lies across a shear-band whose normal is given by a unit vector $\B m$.  In other words, the spatial variation of the displacement occurs along the coordinate $\xi$,
		\begin{equation}
			\xi = \mathbf{m}\cdot \mathbf{x} \ . 
			\label{defxi}
		\end{equation}
		We note that points with the same value of $\xi$ that lie on lines parallel to the band undergo an identical deformation. This reduces a two-dimensional field theory to an effective one-dimensional function along the band normal. For a full derivation of this ansatz the reader is referred to Appendix \ref{ansatz}.
		
		Next we represent the displacement field as
		\begin{equation}
			d_k(\mathbf{x}) = e_k \, f(\xi) .
		\end{equation}
		The unit vector $\mathbf{e}$ is a unit vector orthogonal to $\B m$, while the scalar function $f(\xi)$ specifies the magnitude of the displacement at each position across the band. In the weakly non-linear approach to instabilities the resulting equation is known as an ``amplitude equation" \cite{65Seg,93CH}. Note that this function can be positive or negative. Using this notation, derivatives have the following form
		\begin{equation}
			\partial_{\alpha} = m_{\alpha} \frac{\partial}{\partial \xi}\ ,
		\end{equation}
		And double derivatives can be written as
		\begin{equation}
			\partial_{\alpha} \partial_{\beta}= m_{\alpha} m_{\beta} \frac{\partial ^2}{\partial \xi^2}.
		\end{equation}
		Applying this to $d_k = e_k f(\xi)$, we have
		\begin{equation}
			d_{k, \alpha} = e_k f^{\prime}(\xi) m_{\alpha} \quad  d_{k, \alpha \beta} = e_k f^{\prime \prime}(\xi) m_{\alpha} m_{\beta} \ .
		\end{equation}
		In addition to these relations we will need products of derivatives, which read
		\begin{equation}\label{QP51}
			d_{\mu,\gamma}d_{\mu,\delta} =  {f^{\prime}}^2(\xi)~ m_\gamma m_\delta \ , 
		\end{equation}
		\begin{equation}\label{QP52}
			d_{\mu,\gamma\alpha}d_{\mu,\delta}  =f^{\prime\prime}(\xi) ^{\prime}(\xi) ~m_\gamma m_\alpha m_\delta.
		\end{equation}
		These equations will become handy in analyzing the LHS of Eq.~(\ref{LHSk}).
		
		\subsubsection{Analysis of the LHS}
		
		Let us label the first four terms on the LHS of Eq.~(\ref{LHSk}) as $\mathbb{Z}^{(0)}_k, \mathbb{Z}^{(1)}_k, \mathbb{Z}^{(2)}_k, \mathbb{Z}^{(3)}_k,$ and the final long cubic group as $\mathbb{Z}^{(4)}_k$:
		\begin{widetext}
			\begin{eqnarray}\label{QP47}
				\mathrm{LHS}_k &&= 
				\underbrace{\Sigma^{\alpha\beta} d_{k,\alpha\beta}}_{\mathbb{Z}_k^{(0)}}
				+ \underbrace{\widetilde A_{\alpha k\gamma\delta} d_{\gamma,\delta\alpha}}_{\mathbb{Z}_k^{(1)}}
				+ \underbrace{\frac12 \widetilde A_{\alpha k\gamma\delta}\big(d_{\mu,\gamma\alpha}d_{\mu,\delta}+d_{\mu,\gamma}d_{\mu,\delta\alpha}\big)}_{\mathbb{Z}_k^{(2)}} \\
				&& + \underbrace{\widetilde A_{\alpha\beta\gamma\delta}\big(d_{\gamma,\delta\alpha}d_{k,\beta}+d_{\gamma,\delta}d_{k,\beta\alpha}\big)}_{\mathbb{Z}_k^{(3)}}
				+ \underbrace{\frac12 \widetilde A_{\alpha\beta\gamma\delta}\big(d_{\mu,\gamma\alpha}d_{\mu,\delta}d_{k,\beta}+d_{\mu,\gamma}d_{\mu,\delta\alpha}d_{k,\beta}+d_{\mu,\gamma}d_{\mu,\delta}d_{k,\beta\alpha}\big)}_{\mathbb{Z}_k^{(4)}} \ . \nonumber
			\end{eqnarray}
		\end{widetext}
		At this point we will compute the scalar projection $\Pi$ defined as 
		\begin{equation}\label{QP48}
			\Pi \equiv e_k\,\mathrm{LHS}_k = e_k \mathbb{Z}^{(0)}_k + e_k \mathbb{Z}^{(1)}_k + e_k \mathbb{Z}^{(2)}_k + e_k \mathbb{Z}^{(3)}_k + e_k \mathbb{Z}^{(4)}_k.
		\end{equation}
		We consider each term separately.
		\begin{align}
			\mathbb{Z}_k^{(0)}=e_k \Sigma(m) f''\ , \quad 	e_k \mathbb{Z}_k^{(0)}  =  \Sigma(m) f''
		\end{align}
		where
		\begin{align}
			\Sigma( m)\equiv \Sigma^{\alpha\beta}m_\alpha m_\beta \ . 
			\label{defSig}
		\end{align}
		The second term in Eq.~(\ref{QP47}) reads
		\begin{equation}\label{QP54}
			\mathbb{Z}_k^{(1)}=\widetilde A_{\alpha k\gamma\delta} ~ e_\gamma f''~ m_\delta m_\alpha \ , \quad 	e_k \mathbb{Z}_k^{(1)}=  \mu^{(1)}(m,e)~f''
		\end{equation}
		where the notation $\mu^{(1)}(\hat n,e)$ indicates that it is a projected element of the elastic tensor $\B {\widetilde A}$, 
		
		\begin{equation}\label{QP56}
			\mu^{(1)}(m,e)\equiv e_k~ \widetilde A_{\alpha k\gamma\delta}~ e_\gamma~m_\alpha m_\delta. 
		\end{equation}
		Below we will make this projection explicit using the form of the elastic tensor in homogeneous and isotropic systems. 
		
		The third term in Eq.~(\ref{QP47}) reads
		\begin{equation}\label{QP59}
			\mathbb{Z}^{(2)}_k = f’f^{\prime\prime}\; \widetilde A_{\alpha k\gamma\delta}~m_{\gamma}n_{\delta}m_{\alpha}\ , \quad e_k \mathbb{Z}^{(2)}_k =\mu^{(2)}(m,e) f’f^{\prime\prime}\ , 
		\end{equation}
		where the projection on the elastic tensor here is
		\begin{equation}
			\mu^{(2)}(m,e)\equiv e_k \widetilde A_{\alpha k\gamma\delta}~m_{\gamma}m_{\delta}m_{\alpha} \ .\label{defmu2}
		\end{equation}
		Also this projection will be made explicit below. 
		
		The fourth term in Eq.~(\ref{QP47}) reads
		\begin{eqnarray}\label{QP63}
			&&	\mathbb{Z}^{(3)}_k  = 2\,f’f^{\prime\prime} \; e_k e_\gamma \; \widetilde A_{\alpha\beta\gamma\delta}\; m_{\delta}m_{\beta}m_{\alpha}\ , \nonumber\\ && e_k \mathbb{Z}^{(3)}_k = \,f’f^{\prime\prime} \mu^{(3)}(m,e) \ ,
		\end{eqnarray}
		where in this case the projection on the elastic tensor is 
		\begin{equation}\mu^{(3)}(m,e)\equiv e_\gamma \; \widetilde A_{\alpha\beta\gamma\delta}\; m_{\delta}m_{\beta}m_{\alpha} \ . \label{defmu3}
		\end{equation}
		The last term in Eq.~(\ref{QP47}) reads
		\begin{eqnarray}\label{QP70}
			&&	Z_k^{(4)} = \frac{3}{2} e_k~ f'^2f'' \widetilde A_{\alpha\beta\gamma\delta}~ m_\alpha m_\beta m_\gamma n_\delta\ ,\nonumber \\&& e_k Z_k^{(4)}=\frac{3}{2}~\mu^{(4)}(m)~f'^2f'' \ , 
		\end{eqnarray}
		where
		\begin{align}
			\mu^{(4)}(m) \equiv \widetilde A_{\alpha\beta\gamma\delta}~m_\alpha m_\beta m_\gamma m_\delta \ . \label{defmu4}
		\end{align}
		
		\subsubsection{Simplifying for homogeneous isotropic systems}
		
		Presently we are ready to simplify the result of the previous subsection, taking into account the explicit form of the elastic tensor in homogeneous isotropic systems:
		\begin{equation}
			\widetilde A_{\alpha\beta\gamma\delta}=\lambda~\delta_{\alpha\beta}\delta_{\gamma\delta}
			+\mu(\delta_{\alpha\gamma}\delta_{\beta\delta}+\delta_{\alpha\delta}\delta_{\beta\gamma}) . 
			\label{elten}
		\end{equation}
		Using this in the definitions (\ref{QP56}), (\ref{defmu2}), (\ref{defmu3}) and (\ref{defmu4})
		we find
		\begin{equation}
			\mu^{(1)}=\mu, ~ \mu^{(2)} = \mu^{(3)}=0,  ~\mu^{(4)}=\lambda+2\mu \ .	
		\end{equation}
		
		\subsubsection{Simplifying for simple shear} 
		
		Next we will specialize our equations for the case of simple shear. To this aim we choose the background stress in the form 
		\begin{align}
			\Sigma^{\alpha\beta}= \begin{pmatrix} 0 & \Sigma \\ \Sigma & 0 \end{pmatrix} \ . 
		\end{align}
		At present this tensor is off diagonal in an unspecified coordinate frame that we must make explicit. To this aim we introduce a pair $\B n, \B t$ of orthonormal unit vectors associated with the existent background stress. Fixing the lab frame as the $(x,y)$ coordinates, $\B n \equiv \cos (\theta), \sin (\theta)$ and $\B t\equiv -\sin(\theta), \cos(\theta)$. For any specified plane with a unit normal $\B n$, the traction vector $\B \Sigma \B n$ decomposes uniquely into a normal components $\Sigma(nn) =\B n\cdot \B \Sigma \B n$ and a shear component  $\Sigma(tn)= \B t\cdot \B \Sigma \B n$. These two scalars exhaust the possible traction channels in 2-dimensions. The entire orientation dependence of the traction in simple shear is a point on the Mohr circle defined by
		\begin{equation}
			\left(\Sigma(nn), \Sigma (tn)\right) =\left(\Sigma \sin (2\theta),\Sigma \cos(2\theta)\right) \ .
		\end{equation}
		
		Note that this pair of unit vectors differ from the previously defined pair $\B e, \B m$ that characterize the soft mode orientation. We can express these unit vectors in the basis $(\B n,\B t)$ employing the angle $\phi$ with respect to $\B n$:
		\begin{equation}
			\B m =\B n \cos (\phi)+\B t \sin (\phi) \ ,
		\end{equation}
		with $\B e$ being orthogonal. With this notation we can select any orientation of the soft mode compatible with the background stress.

		Then Eq.~(\ref{defSig}) reads
		\begin{align}
			\Sigma(m) = m_\alpha\Sigma^{\alpha\beta}m_\beta
			= \Sigma (nn) \cos (2\phi)+\Sigma (tn)\sin(2\phi)\ .
		\end{align}
		This is the only place where the background enters in our effective equation. It will be shown to play an important role in driving the instability of the system leading to shear band formation. Note that if $\phi=0$ $\Sigma(m) =\Sigma(nn)$ while if $\phi=45^{0}$ then $\Sigma(m) =\Sigma(tn)$. 
		
		\section{The nonlinear effective equation}
		\label{effective}
		
		Having done with the LHS of Eq.~(\ref{QP26}) we need to perform a similar analysis of its RHS. This is done in detail in Appendix \ref{RHS}. Collecting the results of this appendix 
		with those of the previous section, we can write now the equation for $f(\xi)$, the function that represents the shear band solution:
		\begin{eqnarray}
			\big(\mu+\Sigma(m))\big) f''(\xi) &+& \frac{3}{2}(\lambda+2\mu) f'(\xi)^2f''(\xi) =\nonumber\\& -&\C A f(\xi)+\C B f^3(\xi)\ . \label{beauty}
		\end{eqnarray}
		The parameters $\C A$ and $\C B$ are computed in Appendix \ref{RHS}. 
		To analyze this equation further we will first find the energy functional whose Euler-Lagrange equation agrees with the present equation (\ref{beauty}). 
		\subsection{Energy functional} 
		At this point we can seek an energy functions
		\begin{equation}
			F[f] = \int \mathcal{L}(f, f', \theta)  d \xi
		\end{equation}
		where the Lagrangian density $\mathcal{L}$ is given by:
		\begin{eqnarray}\label{eq:Lagrangian}
			\mathcal{L}(f, f'; \theta)&=& \tfrac{1}{2}\left( \mu + \Sigma(m)\right) (f')^2 + \tfrac{1}{8}(\lambda + 2\mu)(f')^4 \nonumber \\&-& \frac{\mathcal{A}}{2}f^2 + \frac{\mathcal{B}}{4}f^4
		\end{eqnarray}
		The function $f(\xi)$ is supposed to extremize the energy functional, satisfying the Euler-Lagrange equation:
		\begin{equation}\label{EulerLagrange}
			\frac{d}{d \xi}\left( \frac{\partial \mathcal{L}}{\partial f'} \right) - \frac{\partial \mathcal{L}}{\partial f} = 0
		\end{equation}
		We compute the required derivatives from the Lagrangian \eqref{eq:Lagrangian}:
		\begin{align}
			\frac{\partial \mathcal{L}}{\partial f'} &= (\mu + \Sigma(m))f' + \tfrac{1}{2}(\lambda + 2\mu)(f')^3 \\
			\frac{d}{d\xi}\left( \frac{\partial \mathcal{L}}{\partial f'} \right) &= (\mu + \Sigma(m)))f'' + \tfrac{3}{2}(\lambda + 2\mu)(f')^2 f'' \\
			\frac{\partial \mathcal{L}}{\partial f} &= -\mathcal{A}f + \mathcal{B}f^3
		\end{align}
		Using these results in Eq. (\ref{EulerLagrange}) we recapture Eq. (\ref{beauty}), meaning that our energy functional is the right one. 
		\subsection{The Hessian operator}
		
		With the energy functional at hand, we can proceed to compute the Hessian. 
		The Hessian operator $\mathcal{H}$ is defined as the second functional derivative of the energy. For a small variation $\delta f(x)$ around a solution $f(x)$ of \eqref{beauty}, we have:
		
		\begin{equation}
			\delta^2 F = \frac{1}{2} \int \delta f(\xi) \, \mathcal{H} \, \delta f(\xi)~  d \xi
			\label{eq:HessianDef}
		\end{equation}
		To find $\mathcal{H}$, we compute the second variation $\delta^2 F$ directly from the energy functional \eqref{eq:Lagrangian}.
		Consider a variation $f(\xi) \to f(\xi) + \delta f(\xi)$. The first variation $\delta F$ is:
		
		\begin{equation}
			\delta F = \int \left[ \frac{\partial \mathcal{L}}{\partial f} \delta f + \frac{\partial \mathcal{L}}{\partial f'} \delta f' \right] d\xi
		\end{equation}
		Substituting the expressions for the partial derivatives:
		\begin{eqnarray}\label{eq:FirstVariation}
			\delta F &=& \int \Big[ (-\mathcal{A}f + \mathcal{B}f^3)\delta f + \Big\lbrace  (\mu + \Sigma(m))f' \nonumber\\&+& \tfrac{1}{2}(\lambda + 2\mu)(f')^3 \Big\rbrace  \delta f' \Big] d\xi
		\end{eqnarray}
		The second variation $\delta^2 F$ is obtained by varying \eqref{eq:FirstVariation} again, keeping terms up to second order in $\delta f$. 
		\begin{eqnarray}
			&&	\delta[(-\mathcal{A}f + \mathcal{B}f^3)\delta f] =(-\mathcal{A} + 3\mathcal{B}f^2)(\delta f)^2 \\
			&&	\delta[ \Big\lbrace  (\mu + \Sigma(m))f' + \tfrac{1}{2}(\lambda + 2\mu)(f')^3 \Big\rbrace  \delta f' ] = \nonumber \\&&[ \Big\lbrace (\mu + \Sigma(m)) + \tfrac{3}{2}(\lambda + 2\mu)(f')^2 \Big\rbrace ] (\delta f')^2
		\end{eqnarray}
		\\
		Note that the variation of $\delta f'$ is $\delta(\delta f')$, but since $\delta f$ and $\delta f'$ are independent variations, the cross-term $\delta f \, \delta f'$ does not appear in the \textit{second} variation; we only get $(\delta f')^2$.
		Thus, the second variation of the energy functional is:
		
		\begin{eqnarray}
			\delta^2 F &=&\int\Big \{ \Big[ (\mu + \Sigma(m)) + \tfrac{3}{2}(\lambda + 2\mu)(f')^2 \Big] (\delta f')^2 \nonumber\\&+& \Big[ -\mathcal{A} + 3\mathcal{B}f^2 \Big] (\delta f)^2\Big\} d\xi
			\label{eq:SecondVariationRaw}
		\end{eqnarray}
		
		To express $\delta^2 F$ in the form of \eqref{eq:HessianDef}, we integrate the gradient terms by parts, assuming that the boundary terms vanish. 
		For a general function $g(f')$, we have:
		\begin{equation}
			\int g(f') (\delta f')^2  d\xi = - \int \delta f \, \frac{d}{d\xi}\left( g(f') \frac{d}{d\xi} \delta f \right) d\xi
		\end{equation}
		Applying this to our case:
		\begin{widetext}
			\begin{align}
				&\int (\mu + \Sigma(m)) (\delta f')^2  d\xi = - \int \delta f \, (\mu + \Sigma(m)) \frac{d^2}{d\xi^2} ~\delta f(\xi)  ~d\xi \\ \nonumber \\
				&\int \tfrac{3}{2}(\lambda + 2\mu)(f')^2 (\delta f')^2  d\xi = - \int \delta f \, \tfrac{3}{2}(\lambda + 2\mu) \frac{d}{d\xi}\left( (f')^2 \frac{d}{d\xi} \delta f \right) d\xi
			\end{align}
			
			Substituting these results back into \eqref{eq:SecondVariationRaw}:
			
			\begin{align}
				\delta^2 F &= \int \delta f \left\{ -(\mu + \Sigma(m))\frac{d^2}{d\xi^2} - \tfrac{3}{2}(\lambda + 2\mu)\frac{d}{d\xi}\left( (f')^2 \frac{d}{d\xi} \right) \right\} \delta f(\xi)  d\xi 
				&\quad + \int \delta f \left\{ -\mathcal{A} + 3\mathcal{B}f^2 \right\} \delta f(\xi)  d\xi
			\end{align}
			
			Combining both terms, we identify the Hessian operator $\mathcal{H}$ acting on the variation $\delta f$.
			\begin{equation}\label{Hessian}
				\boxed{\mathcal{H} = -\left\lbrace \mu + \Sigma(m)\right\rbrace \frac{d^2}{d\xi^2} - \tfrac{3}{2}(\lambda + 2\mu)\frac{d}{d\xi}\left( (f')^2 \frac{d}{d\xi} \right) - \mathcal{A} + 3\mathcal{B}f^2}
			\end{equation}
		\end{widetext}
		
		In order to guarantee that this Hessian has real eigenvalues and a complete set of orthogonal eigenfunctions we need to show that it is self-adjoint. This is done in Appendix \ref{self}.
		
		\section{The onset of instability}
		\label{onset}
		
		Having the Hessian Eq.~(\ref{Hessian}) at hand, we can ask where an instability sets in. To this aim we remind the reader that a displacement field is always computed with respect to a reference state, and in our quasi-static protocol the last step was the equilibration after the previous strain step. So we need to ask about the stability of a $f=0$ solution. In other words, the point of instability is determined by the Hessian {$\mathcal{H}_{\rm lin}$ of  Eq.~(\ref{Hessian}) in which we substitute $f=f'=0$, i.e.
			\begin{equation}\label{Hlin}
				\mathcal{H}_{\rm lin} = -\left\lbrace \mu + \Sigma(m)\right\rbrace \frac{d^2}{d\xi^2} - \mathcal{A}  \ . 
			\end{equation}
			
			With periodic boundary conditions on a domain of length $L$ the eigenvector of $\mathcal{H}_{\rm lin} $ are Fourier modes,
			\begin{equation}
				\psi_m(\xi)=e^{ik_m\xi},\qquad k_m=\frac{2\pi m}{L},\quad m\in\mathbb Z \ .
			\end{equation}
			and the linear operator (\ref{Hlin}) gives
			\begin{equation}
				\mathcal{H}_{\rm lin} \psi_m=\lambda_m\psi_m
			\end{equation} 
			with
			\begin{equation}
				\lambda_m(\theta)=\Big(\mu+\Sigma(m)\Big)k_m^2-\C A \ . 
			\end{equation}
			The lowest nontrivial mode (typically $m=1$) satisfies
			\begin{equation}
				\lambda_1(\theta)=\Big(\mu+\Sigma(m)\Big)\left(\frac{2\pi}{L}\right)^2-\C A \ .
			\end{equation}
			Instability sets in when 
			\begin{equation}
				\Big(\mu+\Sigma(m)\Big)\left(\frac{2\pi}{L}\right)^2=\C A \ .
				\label{criterion}
			\end{equation}
			
			At this point we should select our boundary conditions, which can be shear traction, $\phi=45^\circ$ or shear normal with $\phi=0$. In the case of simple shear at the boundaries the choice is shear traction. 
			Then  the effect of the background stress is maximal when $\cos(2\theta)=1$  giving $\theta = 0$ or $\theta =90^\circ$. So the well known appearance of the instability in $ 45^\circ $ with respect to the principal shear axis is selected by the physics, not imposed by any geometry or coordinates. So far the formulation is thus internally consistent, coordinate-free, and fully aligned with our physical intuition. 
			
			It is important to realize that the linearized Hessian $	\mathcal{H}_{\rm lin}$ can be only used to identify the point of instability, but it does not provide the solution of the displacement field, a solution that is determined by the full nonlinear equation Eq.~(\ref{beauty}). At this point we are ready to find this solution.

			\section{Shear-band solution of Eq.~(\ref{beauty})}
			\label{tanh}
			
			This section seeks a shear-band solution for the function $f(\xi)$ that satisfies Eq.~(\ref{beauty}). Since we expect such a solution to exhibit all the variation as a function of $\xi$ in the vicinity of the shear band (localized around, say, $\xi_0$), we demand the following boundary conditions
			\begin{eqnarray}
				f(\xi\to-\infty)&=&-f_0 \ , \quad f(\xi\to+\infty) = +f_0
				\nonumber\\ f'(\pm\infty)&=0& \ .  \label{bc}
			\end{eqnarray}
			To find the value of $f_0$ we substitute $f'=0$ in Eq.~(\ref{beauty}) to find 
			\begin{equation}
				f_0=\sqrt{\frac{\C A}{\C B}}	\ . \label{u0}
			\end{equation} 
			At this point it is important to realize that the boundary condition (\ref{u0}) refers to the solution of the projected equation (\ref{beauty}) and not to the full displacement field that is solved by equation (\ref{finalNL}). The latter can include affine displacement fields that are not represented by the projected equation that is purely non-affine.
			
			Armed with the boundary condition we return to Eq.~(\ref{beauty}) and seek a first integral. 
			\subsection{First Integral}
			
			We choose shorthand notation
			\begin{equation}\label{FI2}
				\epsilon_1 = \mu+\Sigma\sin 2\theta,
				\qquad
				\epsilon_2 = \frac{3}{2}(\lambda+2\mu).
			\end{equation}
			Then Eq.~(\ref{beauty}) becomes
			
			\begin{equation}\label{FI3}
				(\epsilon_1+\epsilon_2f'^2)\,f''=-\C A f+\C B f^3.
			\end{equation}
			Multiplying  by $f'(\xi)$:
			\begin{equation}\label{FI4}
				(\epsilon_1+\epsilon_2 f'^2)\,f''f' = (-\C A u+\C B f^3)f'.
			\end{equation}
			Now we use three identities:
			\begin{equation}\label{FI5}
				f''f'=\frac12\frac{d}{d\xi}(f'^2)
			\end{equation}
			
			\begin{equation}\label{FI6}
				f'^2f''f' = f'^3f'' = \frac14\frac{d}{d\xi}(f'^4)
			\end{equation}
			\begin{equation}\label{FI7}
				(-A f+B f^3)f' = \frac{d}{d\xi}\left(-\frac{\C A}{2}f^2+\frac{\C B}{4}f^4\right)
			\end{equation}
			So (\ref{FI4}) becomes a total derivative:
			\begin{equation}\label{FI8}
				\frac{d}{d\xi}\left[
				\frac{\epsilon_1}{2}f'^2+\frac{\epsilon_2}{4}f'^4
				+\frac{\C A}{2}f^2-\frac{\C B}{4}f^4
				\right]=0.
			\end{equation}
			Therefore the first integral is
			\begin{equation}\label{FI9}
				\frac{\epsilon_1}{2}f'^2+\frac{\epsilon_2}{4}f'^4 +\frac{\C A}{2}f^2-\frac{\C B}{4}f^4 = C \ .
			\end{equation}
			Using our boundary conditions Eqs.~(\ref{bc}) and (\ref{u0}) we find
			\begin{equation}\label{FI12}
				C=\frac{\C A}{2}f_0^2-\frac{\C B}{4}f_0^4
				=\frac{\C A^2}{4\C B}.
			\end{equation}
			Substituting back into Eq.(\ref{FI9}) we end up with 
			\begin{equation}\label{FI13}
				\frac{\epsilon_1}{2}f'^2+\frac{\epsilon_2}{4}f'^4=\frac{\C B}{4}(f^2-f_0^2)^2.
			\end{equation}
			
			\subsection{The ductile limit}
			\label{ductile}
			
			In the case of ductile systems we expect the shear-band solution to be relatively smooth, with the slope $f'$ relatively small, allowing us to neglect the quartic term $f'^4$ in Eq.~(\ref{FI13}).
			In other words, we assume that  $\epsilon_1 \gg \epsilon_2 f'^2/2$ in equation (\ref{FI3}); then
			\begin{equation}\label{FI13A}
				\frac{\epsilon_1}{2}f'^2 =\frac{\C B}{4}(f^2-f_0^2)^2.
			\end{equation}
			This equation can be integrated by separation of variables
			\begin{equation}\label{FI14}
				\frac{df}{f_0^2 - f^2} = \sqrt{\frac{\C B}{2\epsilon_1}}\,d\xi\  .
			\end{equation}
			Making use of the standard integral
			\begin{equation}\label{FI15}
				\int \frac{df}{f_0^2 - f^2} = \frac{1}{f_0}\,\mathrm{arctanh}\!\left(\frac{f}{f_0}\right).
			\end{equation}
			one obtains
			\begin{equation}\label{FI16}
				\frac{1}{f_0}\,\mathrm{arctanh}\!\left(\frac{f}{f_0}\right) = \sqrt{\frac{\C B}{2\epsilon_1}}\,(\xi - \xi_0) ,
			\end{equation}
			where $\xi_0$ is the integration constant, identified before as the center of the shear band. Inverting in favor of $f(\xi)$ we find 
			\begin{equation}\label{FI17}
				\frac{f(\xi)}{f_0} = \tanh\!\left(f_0\sqrt{\frac{\C B}{2\epsilon_1}}\,(\xi - \xi_0)\right) .
			\end{equation}
			Since $f_0\sqrt{\C B} = \sqrt{\C A}$, the width is controlled by $\sqrt{\epsilon_1/A}$. We define, 
			\begin{equation}\label{FI18}
				\ell\equiv  \sqrt{\frac{2\epsilon_1}{\C A}} = \sqrt{\frac{2(\mu + \Sigma\sin 2\theta)}{\C A}} .
			\end{equation}
			Then the ductile solution in the final form is:
			\begin{equation}\label{FI19}
				\boxed{
					f(\xi) = f_0\,\tanh\!\left(\frac{\xi - \xi_0}{\ell}\right)} \ ,
			\end{equation}
			with $\ell$ and $f_0$ given by Eqs.~(\ref{FI18}) and (\ref{u0}) respectively. 
			This is the profile obtained directly from the Eq. (\ref{beauty}) in the limit of small  slope.
			At this point we note that in the case that Eq.~(\ref{L1}) survives the nonlinear extension, then
			$\C A=\tilde{\kappa}_e^2$ and the shear band width is directly determined by the screening parameter. 
			
			As and example of how the parameters influence the appearance of the solution (\ref{FI19}) , we plot several cases in Fig.~\ref{ductiled}.  
			\begin{figure}
				\includegraphics[width=0.9 \linewidth]{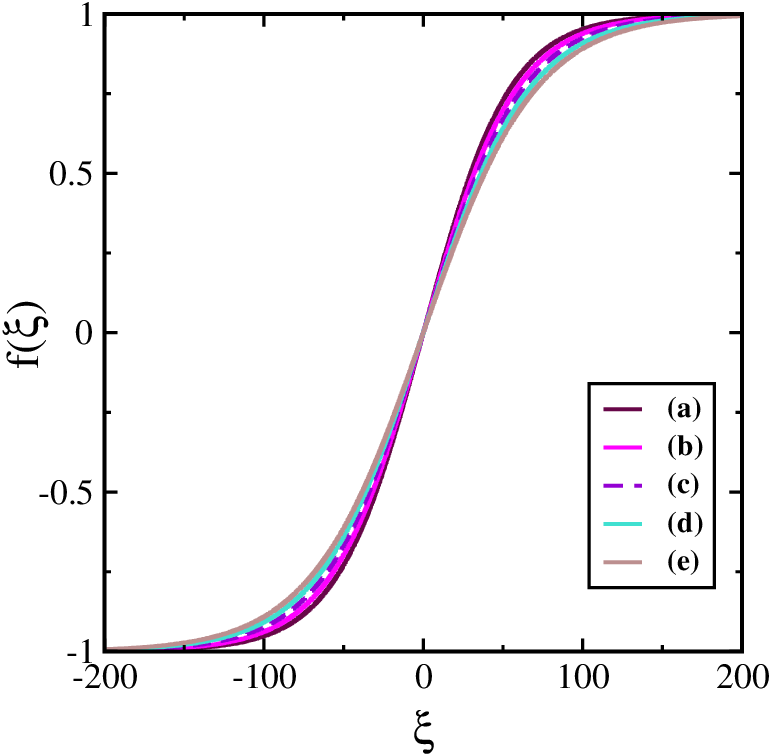}
				\caption{Examples of shear-band profiles in the ductile case. The different plots were obtained with the following parameters: $\theta = 45^\circ$,
					$f_0 = 1$, $\lambda = 50$, $\mu = 2, 4, 6, 8, 10$ (5 cases), $\xi_0 = 0$. This gives characteristic lengths: $\ell = 54, 58, 62, 66, 70$ respectively.}
				\label{ductiled}
			\end{figure}
			The shear associated with these five examples is shown in Fig.~\ref{ductileu}
			\begin{figure}
				\includegraphics[width=0.9 \linewidth]{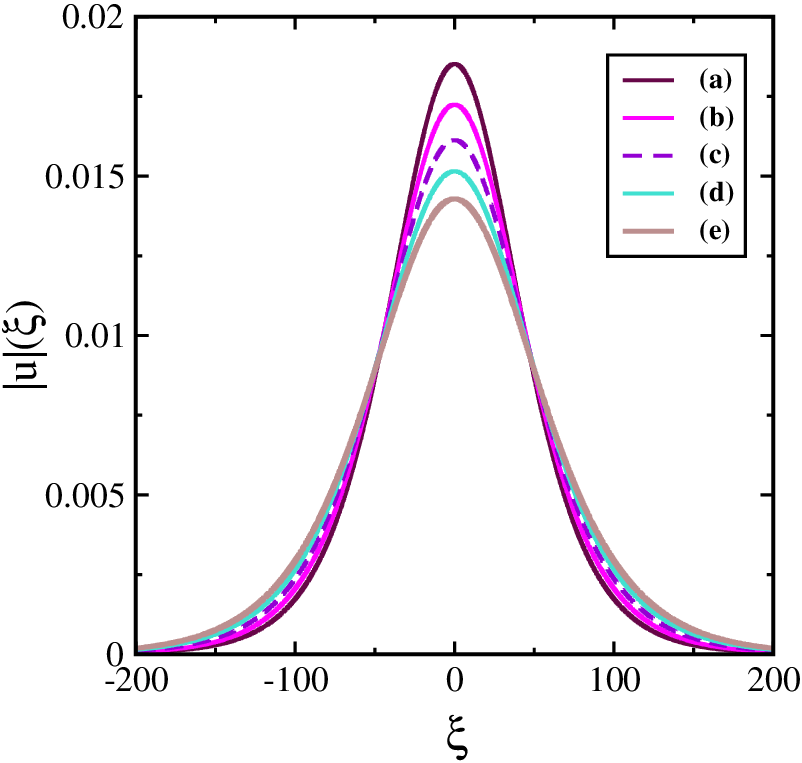}
				\caption{The shear associated with the solution plotted in Fig.~\ref{ductiled}}
				\label{ductileu}
			\end{figure}

			\subsection{The Brittle Limit}
			\label{brittle}
			
			Starting again from Eq.~(\ref{FI13}), we seek a solution in the brittle limit separating the analysis into an inner solution in the region of the shear band, an outer solution for from the shear band, and then matching these two solution. We start with the inner solution.
			\subsubsection{Inner solution}
			
			In the core of the shear band we assume that the quartic term in $f'$ is dominant over the quadratic term, i.e 
			$\epsilon_2(f')^2/2 \gg \epsilon_1$. In that case we can estimate
			\begin{equation}\label{LS5}
				\frac{\alpha}{4}(f')^4 \approx \frac{\C B}{4}(f^2 - f_0^2)^2.
			\end{equation}
			Since shear band solution $f(\xi)$ satisfied $|f(\xi)|\le f_0$, we can take twice a square root of Eq.~(\ref{LS5}) to find
			\begin{equation}\label{LS6}
				f' \approx \left(\frac{\C B}{\epsilon_2}\right)^{1/4} \sqrt{f_0^2 - f^2}.
			\end{equation}
			We integrate this equation by separating variables,
			\begin{equation}\label{LS7}
				\frac{df}{\sqrt{f_0^2 - f^2}} = \left(\frac{B}{\epsilon_2}\right)^{1/4} d\xi \ ,
			\end{equation}
			and after integrating,
			\begin{equation}\label{LS8}
				\arcsin\left(\frac{f}{f_0}\right) = \frac{\xi - \xi_0}{L_c}, \qquad L_c = \left(\frac{\epsilon_2}{\C B}\right)^{1/4} \ .
			\end{equation}
			Thus the solution at the core, which we will denote as $f_{\rm in}(\xi)$ reads
			\begin{equation}\label{LS9}
				f_{\text{in}}(\xi) = f_0 \sin\left(\frac{\xi - \xi_0}{L_c}\right) \ ,
			\end{equation}
			and the core strain is
			\begin{equation}\label{LS10}
				f'_{\text{in}}(\xi) = \frac{f_0}{L_c} \cos\left(\frac{\xi - \xi_0}{L_c}\right) \ .
			\end{equation}
			Next we need to find the outer solution. 
			\subsubsection{The outer solution}
			
			Finding the outer solution is easy, since $f'\to 0$ far from the core, and therefore we face again the condition  $\epsilon_1 \gg \epsilon_2 f'^2$, which is identical to the limit considered in the ductile limit Subsect.~\ref{ductile}. Therefore we can write without further ado an outer solution 
			\begin{equation}\label{LS15}
				f_{\text{out}}(\xi) = f_0 \tanh\left(\frac{\xi - \xi_0}{\ell_0}\right) \ .
			\end{equation}
			\subsubsection{The matched solution}
			Using the results of the last two subsections we can now offer a matched solution,
			\begin{eqnarray}
				f(\xi)&\approx &f_0\sin\left(\frac{\xi-\xi_0}{L_c}\right)\,W(\xi)\nonumber \\&+&f_0\tanh\left(\frac{\xi-\xi_0}{\ell_0}\right)\,[1-W(\xi)] \ ,
			\end{eqnarray}
			where $W(\xi)$ is a smooth function that is $\approx 1$ in the core and $\approx 0$ in the tails.
			
			It is interesting to note that in the brittle limit the width $L_c$ of the shear band is determined by the coefficients appearing in the tensors of the cubic dipole term, rather than the screening parameter $\kappa_e$. Further simulational work will be needed to furnish better understanding of the nature of this typical scale. 
			
			In Appendix \ref{phixi} we show that an alternative form for the solution in the brittle case can be offered in the form $f(\xi) = f_0 \tanh[\phi(\xi)]$ with $\phi(\xi)$ being a nonlinear function of $\xi$. 
			
			To plot the shear band profile and the associated shear profile we need to choose $W$, the smooth interpolation function. We employ the sigmoidal function,
			\begin{equation}
				W(\xi)=\frac{1}{1+\exp\!\left(\frac{|\xi-\xi_0|-\xi_c}{s}\right)}
			\end{equation}
			With this the shear band profile is shown in Fig.~{\ref{brittled} and the shear in Fig.~\ref{brittleu}.
				\begin{figure}
					\includegraphics[width=0.9 \linewidth]{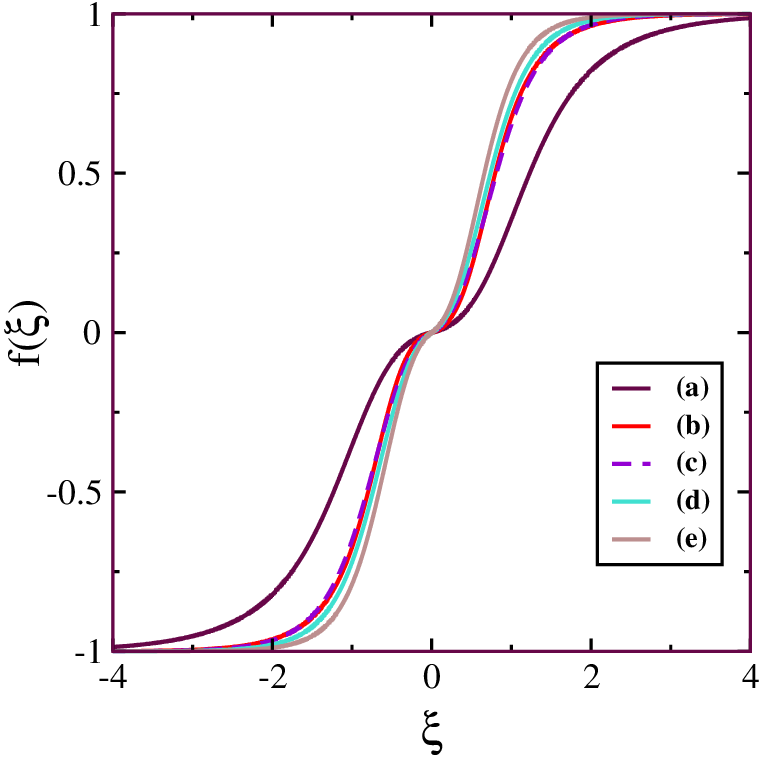}
					\caption{Examples of shear-band profiles in the brittle case. The parameters are shown in the text below}
					\label{brittled}
				\end{figure}
				\begin{figure}
					\includegraphics[width=0.9 \linewidth]{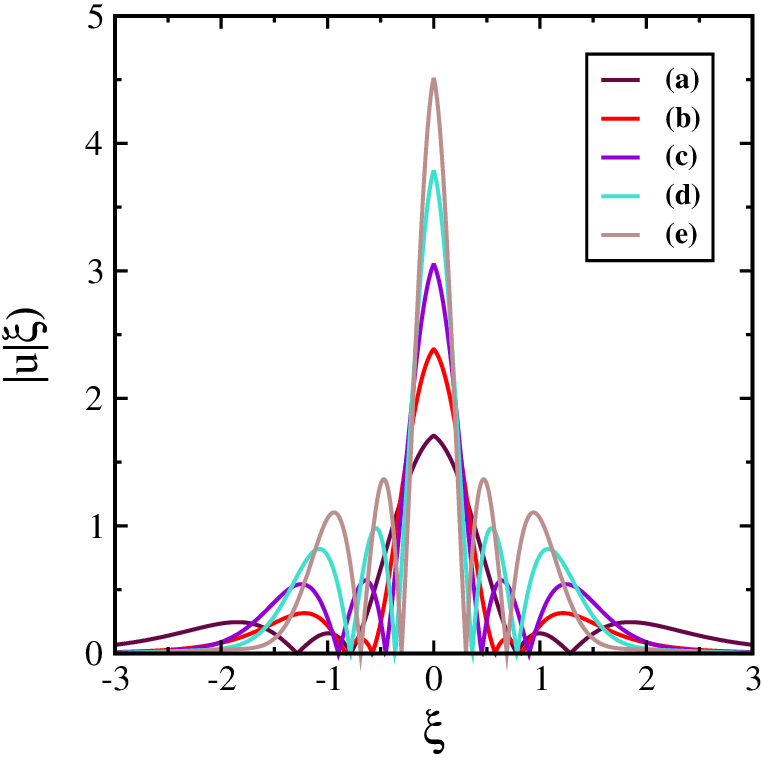}
					\caption{Examples of shear profiles in the brittle case. The parameters are shown in the text below}
					\label{brittleu}
				\end{figure}
				In these two plots the parameters used are $f_0=1$, $\xi_0=0$ and 
				\begin{center}
					\begin{tabular}{|c|c|c|c|c|c|}
						\hline
						Case  & $L_c$ & $\ell_0$ & $\xi_c$ & $s$ \\
						\hline
						(a)  & 0.55 & 1.60 & 0.80 & 0.35 \\
						(b)  & 0.40 & 1.00 & 0.55 & 0.22 \\
						(c)  & 0.30 & 0.95 & 0.55 & 0.28 \\
						(d)  & 0.24 & 0.85 & 0.50 & 0.26 \\
						(e) & 0.20 & 0.75 & 0.45 & 0.24 \\
						\hline
					\end{tabular}
				\end{center}
				
				We note that as expected, decreasing $\ell_0$ and $L_c$ leads to sharper transition in $f(\xi)$.
				Decreasing $s$ leads to a sharper switching region (stronger localization spikes)
				
				\section{Shear localization - zero mode of the nonlinear Hessian}
				\label{sloc}
				
				The Fourier mode which is the eigenfunction of the linearized Hessian served to locate the shear banding instability, but it does not provide any connection to the shear-band solution which results from the nonlinear equation (\ref{beauty}). In this section we show that the zero mode of the nonlinear Hessian is the shear $f'(\xi)$ which is directly exhibiting the shear localization at the shear band. In other words, we show that 
				\begin{equation}\label{H2}
					\mathcal{H}\,\psi_0 = 0
					\quad\text{with}\quad
					\psi_0(\xi) = f'(\xi),
				\end{equation}
				where $f(\xi)$ is the kink solution discussed above. This in fact follows very generally from the translations invariance of the kink solution $f(\xi)$. To see this, we note that
				the energy functional $\mathcal{F}[f] = \int \mathcal{L}(f,f',\theta)\,d\xi$ depends on $\xi$ only through $f(\xi)$, not explicitly on $\xi$; therefore it is invariant under translations. Hence if $f(\xi)$ solves Eq.~(\ref{beauty}), then for any constant shift $\xi_0$, $f_{\xi_0}\equiv f(\xi - \xi_0)$
				also solves Eq.~(\ref{beauty}). Since $f$ is a solution, it satisfies the Euler-Lagrange equation:
				\begin{equation}\label{H4}
					\frac{\delta F}{\delta f}[f] = 0 .
				\end{equation}
				
				But then every translated profile $f_{\xi_0}(\xi)$ is also a solution:
				\begin{equation}\label{H5}
					\frac{\delta F}{\delta f}[f_{\xi_0}] = 0
					\quad\text{for all }\xi_0.
				\end{equation}
				We first note that,
				\begin{equation}\label{H6}
					\frac{\partial f_{\xi_0}(\xi)}{\partial \xi_0} = -\frac{\partial f(\xi - \xi_0)}{\partial \xi} = -f'(\xi - \xi_0).
				\end{equation}
				At $\xi_0 = 0$, we have
				\begin{equation}\label{H7}
					\left. \frac{\partial f_{\xi_0}}{\partial \xi_0} \right|_{\xi_0 = 0} = -f^{\prime}(\xi).
				\end{equation}
				
				Differentiating (\ref{H5}) with respect to $\xi_0$ and then setting $\xi_0 = 0$ (with the chain rule for functionals):
				\begin{equation}\label{H8}
					\left.\frac{d}{d{\xi_0}}\frac{\delta F}{\delta f}[f_{\xi_0}]\right|_{{\xi_0}=0} =0 
				\end{equation}
				\begin{equation}\label{H9}
					\left(\frac{\delta^2 F}{\delta f^2}[f]\right)\left(\left.\frac{d f_{\xi_0}}{d{\xi_0}}\right|_{{\xi_0}=0}\right) = 0.
				\end{equation}
				But
				\begin{equation}\label{H10}
					\left.\frac{d f_{\xi_0}(\xi)}{d{\xi_0}}\right|_{{\xi_0}=0}
					= \left.\frac{d}{d{\xi_0}}\right|_{{\xi_0}=0}f(\xi - {\xi_0})
					= -f'(\xi).
				\end{equation}
				And $\delta^2F/\delta f^2$ is exactly the Hessian operator $H$. So (\ref{H9}) becomes
				\begin{equation}\label{H11}
					\mathcal{H}\,(-f') = 0
					\quad\Rightarrow\quad
					\mathcal{H}\,f' = 0.
				\end{equation}
				\section{Summary and Discussion}
				\label{summary}
				
				It is important to note that the theory presented above contains two types of nonlinearity. The first is the cubic term $\C B u^3(\xi)$ on the RHS of Eq.~(\ref{beauty}). This term acts to tame the linear instability that is analyzed in Sect.~\ref{onset}. Without this nonlinear term the instability will result in a run-away solution that is not saturated. This is the nature for example of the theory of Rudnicki and Rice \cite{75RR,80RR} which was therefore unable to provide a solution of the shear band. The solution that we found above is stabilized by this nonlinearity. 
				Moreover, together with the linear term on the RHS of Eq.~(\ref{beauty}), this term determines how much shear is contained in the shear-banded system, since $f_0 =\sqrt{\C A/\C B}$. 
				
				The second nonlinearity is on the LHS of Eq.~(\ref{beauty}), in the form $3(\lambda+2\mu)\,f'^2\,f''/2$. This nonlinearity controls the \emph{shape} of the band core, not just its existence. This nonlinearity determines \emph{how} shear localizes (ductile vs brittle). When slopes become large, the equation behaves differently: the band core ``feels'' a different stiffness than the far field. This is exactly why we can get a brittle-like modified profile, and not just the weak-slope (ductile) tanh-like profile. This is explained in details in Subsect.~\ref{brittle}. So our theory separates amplitude selection from core-shape selection, and this is quite important. In classical standard $\phi^4$ theories, there is only one intrinsic lengthscale. In our theory, the instability can develop two intrinsic lengthscales in the strong-slope regime (core vs tail), and this is  what brittle localization is expected to show. 
				
				Other attempts to understand shear banding were based on the so called elastoplastic models (including lattice models and mesoscale yielding models) \cite{11VR,16BSMBST,18NFMB,22RBMTZ,22RT}. These models were shown to capture the avalanche statistics and stress redistribution.  However, in their standard form they are local in space, they lack an intrinsic continuum length scale, the localization width is controlled by block sizes or numerical regularization. As a result, shear bands do appear but their thickness is \textit{not} predictive, and their internal strain profile is not uniquely determined. Many Elastoplastic models reproduce the phenomenon of shear localization, but not \textit{how} it localizes spatially in a continuum sense. Our model differs in two decisive ways: it is variational, so equilibrium and stability are well-defined; it contains intrinsic length scales generated by coupling tensors, leading to screened, finite-width shear bands.
				In elastoplastic models the stress redistribution is long-ranged (elastic Green’s function), yielding is local and threshold-based, no energetic penalty exists for sharp strain gradients, no intrinsic screening length appears in the equations. Therefore, localization occurs, but it is not a true screened continuum object. It is a geometrical or numerical localization, not a physically selected one. This is the precise sense in which elastoplastic models are incomplete for predicting a shear-band profile. There are extensions of elastoplastic models that introduce a length scale, such as gradient plasticity, non-local elastoplastic models, the intrinsic length scale is not native to standard elastoplastic models; it must be added explicitly. In our model, it is already built into the Lagrangian.
				
				It should be stressed that without screening, there is no shear band. Plastic rearrangements generate long-range elastic fields (Eshelby-type). If these interactions are unscreened, deformation spreads over the entire system. To obtain localized deformation, these interactions must be screened over a finite length. In our theory, screening arises naturally from coupling parameters in the Lagrangian, the competition between destabilizing and regularizing terms generates a screening length $\ell \sim \sqrt{\frac{\text{gradient stiffness}}{\text{softening strength}}}$. This length directly controls shear-band width.
				A finite shear-band thickness requires an intrinsic material length scale that screens elastic interactions and penalizes sharp strain gradients. Standard elastoplastic models are local and do not possess such an intrinsic length scale; as a result, while they may exhibit localization, the shear-band width is not selected by the physics and depends on discretization or ad-hoc regularization. In contrast, our model contains an intrinsic length scale arising from coupling parameters in the Lagrangian, leading to screened interactions and enabling the emergence of a finite-width, stationary shear-band solution with an explicit core profile. The intimate connection between the instability and screening is underlined by the explicit appearance of the screening parameter $\kappa_e$ in the instability criterion Eq.~(\ref{criterion}). If the linear theory prediction for the tensor $\B K$ continues to be given by Eq.~(\ref{L1}), then $\C A=\tilde{\kappa}_e^2$. Note that only the even component appears in the instability criterion, and that the screening is a stabilizing effect that needs to be overcome by the background stress. 
				
				To proceed, we will need to perform some numerical simulations. First issue for immediate research is the matching of the boundary conditions for the displacement field. We should stress that this should be done with the solution of Eq.~(\ref{finalNL}) rather than with the projected equation (\ref{beauty}). For example, if one has a background affine field that is simple shear, the second derivatives on the LHS of Eq.~(\ref{beauty}) are going to annul it. The shear band solution that we derived is purely non-affine, and needs to be added with correct matching to the affine field to obtain a proper solution of Eq.~(\ref{finalNL}). This is beyond the scope of the present paper, but will be the subject of follow-up research.

				We should also admit that at this point some parameters in our theory are not determined. Although we have expressions for, say, the parameter $\C B$ that determines both the amount of shear carried by the shear band and the brittle inner lengthscale $L_c$, it has never been measured. Thus in the near future we plan to conduct careful numerical simulations to determined the theoretical parameters in classical glass forming models, to test the theory in detail. In light of the considerable progress that has been demonstrated in this paper, we find this future effort quite worthwhile.

				\acknowledgments
				Useful discussions with Gilles Tarjus are thankfully acknowledged. IP acknowledges that this work
				was carried out in part at the Sino-Europe Complex Science
				Center, North University of China, Taiyuan, Shanxi
				030051, China. Thanks to Profs. Guiquan Sun, Xiaofeng
				Luo, for helpful discussions.
				
				\appendix
				\section{Derivation of Eq.~(\ref{LQ13})}
				\label{derive1}
				here we minimize $U$ of Eq.~(\ref{LQ5}) with respect to the quadrupolar field $\B Q$ and displacement field $\B d$ which are the fundamental fields in our problem.
				\begin{align}\label{LQ6}
					\delta_{Q}U &= \int ~d^2x \left( \Lambda_{\alpha\beta\gamma\delta}Q^{\alpha\beta} + \Gamma_{\gamma\delta}^{\alpha\beta}u_{\alpha\beta} \right) \delta Q^{\gamma\delta}, 
				\end{align}
				and
				\begin{align}\label{LQ7}
					\delta_{d}U &= \int d^2x \left(\Sigma^{\alpha\beta} + \sigma^{\alpha\beta}  + \Gamma_{\gamma\delta}^{\alpha\beta} Q^{\gamma\delta}  \right)\delta u_{\alpha\beta} \ .       
				\end{align}
				The variation with respect to $Q$ gives
				\begin{align}\label{LQ8}
					Q^{\alpha\beta} = -\tilde{\Lambda}^{\alpha\beta\gamma\delta}~u_{\gamma\delta}, 
				\end{align}
				where $\tilde{\Lambda}^{\alpha\beta\gamma\delta} = \Lambda^{\alpha\beta\mu\nu}\Gamma_{\mu\nu}^{\alpha\beta}$, and $\Lambda^{\alpha\beta\gamma\delta}$ is the inverse of $\Lambda_{\alpha\beta\gamma\delta}$. 
				Consider now Eqn.~(\ref{LQ7}), and rewrite it as
				\begin{align}\label{LQ9}
					\delta_{d}U &=  
					\int d^2x ~ (\Sigma^{\alpha\beta} + \tilde{\sigma}^{\alpha\beta})~\delta u_{\alpha\beta} \ ,
				\end{align} 
				where $\tilde{\sigma}^{\alpha\beta} = \sigma^{\alpha\beta}  + \Gamma_{\gamma\delta}^{\alpha\beta} Q^{\gamma\delta}$. 
				
				We now employ Eq.~(\ref{NLE21})
				and do the integration by parts, to get the following equation
				\begin{eqnarray}\label{LQ11}
					\delta_d U  &= &\left[ \int_{\partial \Omega}  \tilde{\sigma}_{\alpha\beta} ~\hat{n}_{\beta}~ \delta d_{\alpha}  ~ds +  \int_{\partial \Omega} \tilde{\sigma}_{\alpha\beta} d_{k,\alpha}\hat{n}_{\beta}~ \delta d_{k} ~ds    \right]\nonumber \\ &-& \left[  \int d^2x \left[  \tilde{\sigma}_{k\beta, \beta}   +   (\tilde{\sigma}_{\alpha\beta}d_{k,\alpha})_{, \beta} \right]~ \delta d_{k}  \right]  \nonumber \\
					& + &\left[ \int_{\partial \Omega}  \Sigma^{\alpha\beta} ~\hat{n}_{\beta}~ \delta d_{\alpha}  ~ds +  \int_{\partial \Omega} \Sigma^{\alpha\beta} d_{k,\alpha}\hat{n}_{\beta}~ \delta d_{k} ~ds    \right] \nonumber\\&-& \left[  \int d^2x \left[ {\Sigma^{k\beta}_{, \beta}}   +   (\Sigma^{\alpha\beta}d_{k,\alpha})_{, \beta} \right]~ \delta d_{k} \right]=0.
				\end{eqnarray}
				which gives the equilibrium equation
				\begin{align}\label{ApLQ13}
					(\Sigma^{\alpha\beta}d_{k,\beta})_{, \alpha} + \tilde{\sigma}_{\alpha k, \alpha}   +   (\tilde{\sigma}_{\alpha \beta}d_{k,\beta})_{, \alpha} =0.
				\end{align}

				
				\section{Energy-Functional for Equilibrium Equation, Its Hessian, Instability and Soft-Mode}
				\label{ansatz}
				
				The aim of this appendix is to justify from first principles the projection procedure employed in Sect.~\ref{anal} and to establish its connection to the instability analysis performed in Sect.~\ref{onset}. 
				
				We note that the equilibrium equation, which is a vector equation, is written for the displacement field $d$. In this section, we will demonstrate that the vector Equilibrium equation can be derived from an energy functional that is explicitly written in terms of the displacement field $d$. 
				From this energy functional, we compute the Hessian, which is the second variation of the energy functional, and show that it is a self-adjoint operator. To study the instability, we linearize the Hessian around the homogeneous state $d=0$. 
				
				Since the Hessian is a self-adjoint operator near the instability, we write the displacement field as an expansion in the orthonormal eigenmodes. We then demonstrate that the zero mode, corresponding to the instability, is the dominant mode. Therefore, at the instability leading to shear banding, it is appropriate to approximate the displacement field by the zero mode of the Hessian.
				
				Then, we write the eigenvalue equation from this Hessian operator and compute its eigenvalues and eigenvectors, which represent its polarization. From these polarizations, we show that transverse polarization facilitates the shear localization. Finally, we write down the structure of the soft mode for the displacement field, which is our projection ansatz leading to post-instability shear band.

				\subsection{The unprojected energy functional $U[d]$}
				We note that the governing equilibrium equation is an Euler-Lagrange equation so it should be derived form the variation of anenergy functional $U[d]$, that is,
				$
				\delta U[d]=0$ for all admissible variations $\eta$ satisfying  $\eta|_{\partial\Omega}=0$, 
				such that after integration by parts we get a PDE of the form (\ref{beauty}).
				
				Guided by Eq.~\ref{beauty}, we write the energy functional $U[d]$ in the form
				\begin{widetext}
					\begin{equation}
						U[d]=\int_\Omega {\cal W} d^2x\equiv \int_\Omega\Bigg[~\frac12\,\Sigma^{\alpha\beta}\,d_{k,\alpha}d_{k,\beta}+\frac12\,\tilde A_{\alpha\beta\gamma\delta}\,u_{\alpha\beta}(d)\,u_{\gamma\delta}(d)+ \frac12\,L^{k\zeta}d_k d_\zeta-\frac14\,T^{kpqr}d_k d_p d_q d_r\Bigg]{d^2 x}. 
					\end{equation}
				\end{widetext} 
				
				As a shorthand representation for further analysis we rewrite this functional as 
				\begin{equation}
					U[d]=\int_\Omega \left[ W_{\infty}(\nabla d)+W_{\text{el}}(d,\nabla d)+W_{\text{loc}}(d)\right]\,{d^2 x},  \label{Uofd}.
				\end{equation}
				with\\
				(i) Constant prestress-gradient energy density
				
				\begin{equation}
					W_{\infty}=\frac12\,\Sigma^{\alpha\beta}\,d_{k,\alpha}\,d_{k,\beta}.
					\label{E1}
				\end{equation}

				(ii) Elastic energy density (built from the finite strain)
				
				\begin{equation}
					W_{\text{el}}=\frac12\,\tilde A_{\alpha\beta\gamma\delta}\,u_{\alpha\beta}(d)\,u_{\gamma\delta}(d). \label{E2}
				\end{equation}

				(iii) Local Landau-type potential 
				
				\begin{equation}
					W_{\text{loc}} =\frac12\,L_{k\zeta}d_k d_\zeta -\frac14\,T_{kpqr}d_k d_p d_q d_r.  \label{E3}
				\end{equation}
				
				The Euler-Lagrange equation is:
				\begin{equation}
					-\frac{\delta U}{\delta d_k}= -\frac{\partial \mathcal W}{\partial d_k}+\partial_\alpha\!\left(\frac{\partial \mathcal W}{\partial d_{k,\alpha}}\right).
				\end{equation}
				We will compute this term by term.
				\begin{equation}
					-\frac{\partial W_{\text{loc}}}{\partial d_k}= -L^{k\zeta}d_\zeta + T^{kpqr}d_p d_q d_r.
				\end{equation}
				This already matches the RHS of equilibrium equation.
				
				\begin{equation}
					\frac{\partial W_{\rm el}}{\partial d_{k,\alpha}}= \sigma_{\alpha k}+\sigma_{\alpha\beta}d_{k,\beta}.
				\end{equation}
				
				Therefore,
				
				\begin{equation}
					\frac{\partial \mathcal W}{\partial d_{k,\alpha}}= \Sigma^{\alpha\beta}d_{k,\beta}+\sigma_{\alpha k}+\sigma_{\alpha\beta}d_{k,\beta}. 
				\end{equation}
				
				Thus, 
				
				\begin{equation}
					\partial_\alpha\!\left(\frac{\partial \mathcal W}{\partial d_{k,\alpha}}\right)= 
					\big(\Sigma^{\alpha\beta}d_{k,\beta}\big)_{,\alpha} +\sigma_{\alpha k,\alpha} +\big(\sigma_{\alpha\beta}d_{k,\beta}\big)_{,\alpha}. 
				\end{equation}
				
				In summary we write
				\begin{multline}
					-\frac{\delta U}{\delta d_k}= 
					\big(\Sigma^{\alpha\beta}d_{k,\beta}\big)_{,\alpha}+\sigma_{\alpha k,\alpha}+\big(\sigma_{\alpha\beta}d_{k,\beta}\big)_{,\alpha}\\
					-\,L^{k\zeta}d_\zeta +\,T^{kpqr}d_p d_q d_r.  
				\end{multline}
				\\
				
				Setting $(-\delta U/\delta d_k=0)$ gives us exactly the equilibrium equation.

				\subsection{Full unprojected Hessian from the Energy Functional $U[d]$}
				
				Let us take two arbitrary perturbations $\eta,\zeta$. We write the Hessian which is defined as the bilinear second variation of $U$,
				\begin{equation}
					\delta^2U[d;\eta,\zeta]=\left.\frac{\partial^2}{\partial s\,\partial t}\right|_{s=t=0}U[d+s\eta+t\zeta] ,
				\end{equation}
				together with periodic boundary conditions, or $d$ fixed on $\partial\Omega$ so that the admissible variations satisfy
				\begin{equation}
					\eta_k|_{\partial\Omega}=0, \qquad \zeta_k|_{\partial\Omega}=0. 
				\end{equation}
				
				\subsubsection{First variation: computing $\delta U[d;\eta]$}
				
				Let $d \mapsto d+\varepsilon\eta$. We define the first variation,
				
				\begin{equation}
					\delta U[d;\eta]=\left.\frac{d}{d\varepsilon}\right|_{\varepsilon=0}U[d+\varepsilon\eta].
				\end{equation}
				
				Now we compute each term separately,
				\begin{equation}
					U_\infty[d]=\int_\Omega \frac12\Sigma^{\alpha\beta}d_{k,\alpha}d_{k,\beta}\,{d^2x}. \nonumber
				\end{equation}
				
				Differentiate:
				
				\begin{equation}
					\delta U_\infty[d;\eta]=
					\int_\Omega \frac12\Sigma^{\alpha\beta}\Big(\eta_{k,\alpha}d_{k,\beta}+d_{k,\alpha}\eta_{k,\beta}\Big)\,{d^2 x}. \nonumber
				\end{equation}
				
				If $\Sigma^{\alpha\beta}=\Sigma^{\beta\alpha}$ (standard Cauchy stress), the two terms are equal, giving
				
				\begin{equation}\label{dUinf}
					\delta U_\infty[d;\eta] =\int_\Omega \Sigma^{\alpha\beta}d_{k,\beta}\eta_{k,\alpha}\,{d^2 x}. 
				\end{equation}
				
				The elastic energy is:
				
				\begin{equation}
					U_{\rm el}[d]=\int_\Omega \frac12\tilde A_{\alpha\beta\gamma\delta}u_{\alpha\beta}(d)u_{\gamma\delta}(d)\,{d^2 x}. \nonumber
				\end{equation}
				
				Differentiate:
				
				\begin{align}\label{Uel}
					\delta U_{\rm el}[d;\eta] &=\int_\Omega \tilde A_{\alpha\beta\gamma\delta}u_{\gamma\delta}(d)\delta u_{\alpha\beta}[d;\eta]\,{d^2x} \nonumber \\
					&=\int_\Omega \sigma_{\alpha\beta}(d)\delta u_{\alpha\beta}[d;\eta]\,{d^2 x}.
				\end{align}
				
				So we must compute $\delta u_{\alpha\beta}$ explicitly from (\ref{defunl}):
				Taking the variation term by term:
				\begin{align}
					\delta(d_{\alpha,\beta})&=\eta_{\alpha,\beta},\nonumber \\
					\delta(d_{\beta,\alpha})&=\eta_{\beta,\alpha}, \nonumber \\
					\delta(d_{\mu,\alpha}d_{\mu,\beta})&=\eta_{\mu,\alpha}d_{\mu,\beta}+d_{\mu,\alpha}\eta_{\mu,\beta}. \nonumber
				\end{align}
				Therefore
				
				\begin{equation}\label{uab}
					\delta u_{\alpha\beta}[d;\eta]=\frac12\Big(\eta_{\alpha,\beta}+\eta_{\beta,\alpha}+\eta_{\mu,\alpha}d_{\mu,\beta}+d_{\mu,\alpha}\eta_{\mu,\beta}\Big).
				\end{equation}
				
				\begin{equation}
					U_L[d]=\int_\Omega \frac12 L^{k\zeta}d_k d_\zeta\,{d^2 x}. \nonumber
				\end{equation}
				
				Then
				
				\begin{equation}\label{L}
					\delta U_L[d;\eta]=\int_\Omega L^{k\zeta}d_\zeta\eta_k\,{d^2x},\qquad\left(\frac{\delta U}{\delta d_k}\right)_L = L^{k\zeta}d_\zeta.
				\end{equation}
				
				\begin{equation}
					U_T[d]=\int_\Omega -\frac14 T^{kpqr}d_k d_p d_q d_r\,{d^2x}. \nonumber
				\end{equation}
				
				Let us write the first variation first. The variation of each $d$ is computed. Differentiating with respect to $d_k,~d_p,~d_q$ and $d_r$ gives $-\frac14 T^{kpqr} \eta_k d_p d_q d_r$, \(-\frac14 T^{kpqr} d_k \eta_p d_q d_r\), \(-\frac14 T^{kpqr} d_k d_p \eta_q d_r\), and \(-\frac14 T^{kpqr} d_k d_p d_q \eta_r\) respectively.  Adding all the terms with the symmetry of $T^{kpqr}$ in $(k, p,q,r)$, the result is
				
				\begin{multline}\label{T}
					\delta U_T[d;\eta]=\int_\Omega -T^{kpqr}d_p d_q d_r\eta_k\,{d^2x}, \\ \left(\frac{\delta U}{\delta d_k}\right)_T = -T^{kpqr}d_p d_q d_r. 
				\end{multline}

				\subsection{Second variation: Full unprojected Hessian in bilinear form}
				
				\subsubsection{Prestress-gradient term}
				
				From (\ref{dUinf}):
				
				\begin{equation}
					\delta U_\infty[d;\eta]=\int \Sigma^{\alpha\beta}d_{k,\beta}\eta_{k,\alpha}. \nonumber
				\end{equation}
				
				We vary this with respect to $d$ in direction $\zeta$: $d_{k,\beta}\mapsto d_{k,\beta}+t\zeta_{k,\beta}$. Thus
				
				\begin{equation}\label{d2Uinf}
					\delta^2U_\infty[d;\eta,\zeta] =\int_\Omega \Sigma^{\alpha\beta}\zeta_{k,\beta}\eta_{k,\alpha}\,{d^2x}.
				\end{equation}
				
				This is already symmetric in $(\eta,\zeta)$ if $\Sigma^{\alpha\beta}$ is symmetric (interchanging dummy indices $\alpha\leftrightarrow\beta$).
				
				\subsubsection{Local terms}
				
				Quadratic term (\ref{L}):
				
				\begin{equation}
					\delta^2U_L[d;\eta,\zeta]=\int_\Omega \eta_k L^{k\zeta}\zeta_\zeta\,{d^2x}.
				\end{equation}
				
				Quartic term: from (\ref{T}), vary $-T^{kpqr}d_p d_q d_r$ w.r.t. \(d\) in direction $\zeta$. Derivatives with respect to $d_p$,  $d_q$, and $d_r$: gives $-T^{kpqr} \zeta_p d_q d_r $, $-T^{kpqr} d_p \zeta_q d_r$, and $-T^{kpqr} d_p d_q \zeta_r$ respectively. Adding all the three terms with symmetry in $(p,q,r)$ we get
				
				\begin{equation}
					\delta^2U_T[d;\eta,\zeta]=\int_\Omega -3\eta_k T^{k\zeta qr}d_q d_r\zeta_\zeta\,{d^2x}.
				\end{equation}
				
				So the combined ``local stiffness'' tensor is
				
				\begin{align}\label{M}
					M^{k\zeta}(d) & =L^{k\zeta}-3T^{k\zeta qr}d_q d_r, \nonumber \\
					\delta^2U_{\rm loc}[d;\eta,\zeta]& =\int \eta_k M^{k\zeta}(d)\zeta_\zeta\,{d^2x}.
				\end{align}
				
				\subsubsection{Elastic term}
				
				From equation (\ref{Uel}):
				
				\begin{equation}
					\delta U_{\rm el}[d;\eta]=\int \sigma_{\alpha\beta}(d)\delta u_{\alpha\beta}[d;\eta]\,{d^2x}. \nonumber
				\end{equation}
				
				Now we take variation w.r.t. $d$ in direction $\zeta$. There are two places where $d$ appears:
				
				\begin{itemize}
					\item inside $\sigma_{\alpha\beta}(d)$,
					\item inside $\delta u_{\alpha\beta}[d;\eta]$ (because (\ref{uab}) depends on $d$ via $d_{\mu,\alpha}$ terms).
				\end{itemize}
				
				So:
				\begin{multline}\label{UEL}
					\delta^2U_{\rm el}[d;\eta,\zeta]= \int (\delta\sigma_{\alpha\beta}[d;\zeta])\delta u_{\alpha\beta}[d;\eta]\,{d^2x} \\
					+ \int \sigma_{\alpha\beta}(d)\delta(\delta u_{\alpha\beta}[d;\eta])\,{d^2x}.
				\end{multline}
				
				\subsubsection{Step 1: compute $\delta\sigma_{\alpha\beta}[d;\zeta]$}
				
				From $\sigma_{\alpha\beta}=\tilde A_{\alpha\beta\gamma\delta}u_{\gamma\delta}$,
				
				\begin{equation}\label{ds}
					\delta\sigma_{\alpha\beta}[d;\zeta]=\tilde A_{\alpha\beta\gamma\delta}\delta u_{\gamma\delta}[d;\zeta].
				\end{equation}
				
				and $\delta u_{\gamma\delta}[d;\zeta]$ is given explicitly by (\ref{uab}) with $\eta\to\zeta$.
				
				\subsubsection{Step 2: compute $\delta(\delta u_{\alpha\beta}[d;\eta])$}
				
				From (\ref{uab}),
				
				\begin{equation}
					\delta u_{\alpha\beta}[d;\eta]= \frac12\Big(\eta_{\alpha,\beta}+\eta_{\beta,\alpha}
					+\eta_{\mu,\alpha}d_{\mu,\beta}+d_{\mu,\alpha}\eta_{\mu,\beta}\Big). \nonumber
				\end{equation}
				
				Only the last two terms depend on $d$. Vary $d_{\mu,\beta}\mapsto d_{\mu,\beta}+t\zeta_{\mu,\beta}$, $d_{\mu,\alpha}\mapsto d_{\mu,\alpha}+t\zeta_{\mu,\alpha}$. Then
				
				\begin{multline}\label{d2u}
					\delta(\delta u_{\alpha\beta}[d;\eta])\ \text{in direction }\zeta=
					\frac12\Big(\eta_{\mu,\alpha}\zeta_{\mu,\beta}+\zeta_{\mu,\alpha}\eta_{\mu,\beta}\Big) \\ \equiv \delta^2u_{\alpha\beta}[d;\eta,\zeta].
				\end{multline}

				\subsubsection{Insert into (\ref{UEL})}
				
				Using (\ref{ds}) and (\ref{d2u}):
				
				\begin{multline}
					\delta^2U_{\rm el}[d;\eta,\zeta]=
					\int_\Omega
					\tilde A_{\alpha\beta\gamma\delta}
					\delta u_{\gamma\delta}[d;\zeta]
					\delta u_{\alpha\beta}[d;\eta]\,{d^2x} \\
					+
					\int_\Omega
					\sigma_{\alpha\beta}(d)
					\delta^2u_{\alpha\beta}[d;\eta,\zeta]\,{d^2x}.
				\end{multline}
				
				Now we rewrite the first integral symmetrically by swapping the dummy $(\alpha\beta)\leftrightarrow(\gamma\delta)$ and using the major symmetry $\tilde A_{\alpha\beta\gamma\delta}=\tilde A_{\gamma\delta\alpha\beta}$. Now it becomes symmetric in $(\eta,\zeta)$. Therefore, the elastic bilinear form is:
				
				\begin{multline}\label{d2Uel}
					\delta^2U_{\rm el}[d;\eta,\zeta]=
					\int_\Omega
					\tilde A_{\alpha\beta\gamma\delta}
					\delta u_{\alpha\beta}[d;\eta]
					\delta u_{\gamma\delta}[d;\zeta]\,{d^2x} \\
					+
					\int_\Omega
					\sigma_{\alpha\beta}(d)
					\frac12\Big(\eta_{\mu,\alpha}\zeta_{\mu,\beta}+\zeta_{\mu,\alpha}\eta_{\mu,\beta}\Big)\,{d^2x}.
				\end{multline}
				
				Everything here is explicit because $\delta u$ is (\ref{uab}).
				
				\subsection{Full unprojected Hessian in bilinear form}
				
				Collect (\ref{d2Uinf}), (\ref{M}), and (\ref{d2Uel}):
				
				\begin{multline}
					\delta^2U[d;\eta,\zeta]=\int_\Omega \Sigma^{\alpha\beta}\eta_{k,\alpha}\zeta_{k,\beta}\,{d^2x}
					+\delta^2U_{\rm el}[d;\eta,\zeta]\\
					+ \int_\Omega \eta_k M^{k\zeta}(d)\zeta_\zeta\,{d^2x},
				\end{multline}
				
				with $M^{k\zeta}(d)=L^{k\zeta}-3T^{k\zeta qr}d_q d_r$ and $\delta^2U_{\rm el}$ given by (\ref{d2Uel}). This is the full unprojected Hessian.
				
				\subsection{Writing to operator form $\mathcal H_U[d]$ explicitly}
				
				Now we define $\mathcal H_U[d]$ such that,
				
				\begin{equation}\label{HUd}
					\delta^2U[d;\eta,\zeta]=\int_\Omega \eta_k(\mathcal H_U[d]\zeta)_k\,{d^2x} \quad\text{for all }\eta.
				\end{equation}
				
				We must integrate by parts in each term to move derivatives off $\eta$. Now let us compute the operator term-by-term.
				
				\subsubsection{Prestress operator part}
				
				Start from (\ref{d2Uinf}):
				
				\begin{equation}
					\int \Sigma^{\alpha\beta}\eta_{k,\alpha}\zeta_{k,\beta}
					=
					-\int \eta_k(\Sigma^{\alpha\beta}\zeta_{k,\beta})_{,\alpha}\,{d^2x}. \nonumber
				\end{equation}
				
				So
				
				\begin{equation}\label{Hinf}
					(\mathcal H_\infty\zeta)_k=-(\Sigma^{\alpha\beta}\zeta_{k,\beta})_{,\alpha}.
				\end{equation}
				
				\subsubsection{Local operator part}
				
				From (\ref{M}):
				
				\begin{equation}\label{Hloc}
					\int \eta_k M^{k\zeta}(d)\zeta_\zeta\,{d^2x}
					\Rightarrow
					(\mathcal H_{\rm loc}[d]\zeta)_k=M^{k\zeta}(d)\zeta_\zeta.
				\end{equation}
				
				\subsection{Elastic operator part}
				
				Take the elastic bilinear form (\ref{d2Uel}) and rewrite it as $\int \eta_k(\cdots)\,{d^2x}$.
				
				\subsubsection*{(i) ``Material'' elastic part}
				
				\begin{equation}
					I_1:=\int \tilde A_{\alpha\beta\gamma\delta}
					\delta u_{\alpha\beta}[d;\eta]
					\delta u_{\gamma\delta}[d;\zeta]\,{d^2x}.
				\end{equation}
				
				Insert explicit $\delta u_{\alpha\beta}[d;\eta]$ from (\ref{uab}):
				
				\begin{equation}
					\delta u_{\alpha\beta}[d;\eta]
					=\frac12\Big(\eta_{\alpha,\beta}+\eta_{\beta,\alpha}+\eta_{\mu,\alpha}d_{\mu,\beta}+d_{\mu,\alpha}\eta_{\mu,\beta}\Big). \nonumber
				\end{equation}
				
				Therefore $I_1$ is a sum of terms each linear in $\eta_{k,\alpha}$. After collecting each them, we obtain:
				
				\begin{equation}
					I_1=\int (\delta\sigma_{\alpha k}[d;\zeta]+\delta\sigma_{\alpha\beta}[d;\zeta]d_{k,\beta})\eta_{k,\alpha}\,{d^2x}.
				\end{equation}
				
				Where, by definition,
				
				\begin{equation}
					\delta\sigma_{\alpha\beta}[d;\zeta]=\tilde A_{\alpha\beta\gamma\delta}\delta u_{\gamma\delta}[d;\zeta]. \nonumber
				\end{equation}
				
				So $\delta\sigma$ is ``$\tilde A \delta u(\zeta)$''. We can understand the derivation of $I_1$ in simple steps also. Note that the first variation of $U[d]$ is given by:
				
				\begin{equation}
					\delta U_{\rm el}[d;\eta]=\int(\sigma_{\alpha k}+\sigma_{\alpha\beta}d_{k,\beta})\eta_{k,\alpha}\,{d^2x}. \nonumber
				\end{equation}
				
				Now we replace $\sigma$ by $\delta\sigma[d;\zeta]$, this is because $I_1$ is exactly ``variation of the elastic first variation in the stress factor'', and we get ($I_1$). After integrating by parts, $I_1$ is reduced to
				
				\begin{equation}
					I_1= -\int \eta_k\Big[(\delta\sigma_{\alpha k}[d;\zeta])_{,\alpha}
					+(\delta\sigma_{\alpha\beta}[d;\zeta]d_{k,\beta})_{,\alpha}\Big]{d^2x}.
				\end{equation}
				
				So the contribution from ``material elastic operator'' is
				
				\begin{equation}
					(\mathcal H_{\rm el}^{(1)}[d]\zeta)_k =
					-(\delta\sigma_{\alpha k}[d;\zeta])_{,\alpha}
					-(\delta\sigma_{\alpha\beta}[d;\zeta]d_{k,\beta})_{,\alpha}.
				\end{equation}
				
				\subsubsection{(ii) “Geometric” elastic part}
				
				Next, we compute the second term in (\ref{d2Uel}):
				
				\begin{equation}
					I_2:=\int \sigma_{\alpha\beta}(d)\frac12(\eta_{\mu,\alpha}\zeta_{\mu,\beta}+\zeta_{\mu,\alpha}\eta_{\mu,\beta})\,{d^2x}.
				\end{equation}
				Using symmetry $\sigma_{\alpha\beta}=\sigma_{\beta\alpha}$, giving
				\begin{equation}
					I_2=\int \sigma_{\alpha\beta}(d)\eta_{\mu,\alpha}\zeta_{\mu,\beta}\,{d^2x}.
				\end{equation}
				
				relabelling $\mu\to k$:
				\begin{equation}
					I_2=\int \sigma_{\alpha\beta}(d)\eta_{k,\alpha}\zeta_{k,\beta}\,{d^2x}.
				\end{equation}
				Integrating by parts in $\alpha$:
				\begin{equation}
					I_2=-\int \eta_k(\sigma_{\alpha\beta}(d)\zeta_{k,\beta})_{,\alpha}\,{d^2x}.
				\end{equation}
				So the geometric elastic operator is
				\begin{equation}
					(\mathcal H_{\rm el}^{(2)}[d]\zeta)_k=-(\sigma_{\alpha\beta}(d)\zeta_{k,\beta})_{,\alpha}.
				\end{equation}
				
				\textit{Now combining elastic operator parts:}
				\begin{multline}\label{Hel}
					(\mathcal H_{\rm el}[d]\zeta)_k=
					-(\delta\sigma_{\alpha k}[d;\zeta])_{,\alpha}
					-(\delta\sigma_{\alpha\beta}[d;\zeta]d_{k,\beta})_{,\alpha} \\
					-(\sigma_{\alpha\beta}(d)\zeta_{k,\beta})_{,\alpha}.
				\end{multline}
				with:
				\begin{multline}
					\delta\sigma_{\alpha\beta}[d;\zeta]
					=\tilde A_{\alpha\beta\gamma\delta}\delta u_{\gamma\delta}[d;\zeta],~~\text{where}~~ \delta u_{\gamma\delta}[d;\zeta]\\
					\text{is given by (\ref{uab}) with }\eta\to\zeta.
				\end{multline}
				
				\subsubsection{Full unprojected Hessian operator}
				
				Combine (\ref{Hinf}), (\ref{Hloc}), (\ref{Hel}):
				
				\begin{widetext}
					\begin{equation}
						(\mathcal H_U[d]\zeta)_k =
						-(\Sigma^{\alpha\beta}\zeta_{k,\beta})_{,\alpha}
						-(\delta\sigma_{\alpha k}[d;\zeta])_{,\alpha}
						-(\delta\sigma_{\alpha\beta}[d;\zeta]d_{k,\beta})_{,\alpha}
						-(\sigma_{\alpha\beta}(d)\zeta_{k,\beta})_{,\alpha}
						+ M^{k\zeta}(d)\zeta_\zeta,
					\end{equation}
				\end{widetext}
				
				with $M^{k\zeta}(d)=L^{k\zeta}-3T^{k\zeta qr}d_q d_r$. This is the full unprojected nonlinear Hessian operator derived from the energy.

				\subsection{Self-adjointness}
				
				We use the $L^2$ inner product
				
				\begin{equation}
					\langle a,b\rangle = \int_\Omega a_k b_k\,{d^2x}.
				\end{equation}
				
				Self-adjointness of $\mathcal H_U[d]$ means:
				
				\begin{equation}\label{SA}
					\langle \eta,\mathcal H_U[d]\zeta\rangle= \langle \zeta,\mathcal H_U[d]\eta\rangle \quad\text{for all admissible }\eta,\zeta.
				\end{equation}
				
				\subsubsection{Symmetry of the second variation}
				
				Because $U[d]$ is a scalar functional in $d$, its mixed second derivatives commute:
				
				\begin{multline}\label{D2U}
					\delta^2U[d;\eta,\zeta]
					=
					\left.\frac{\partial^2}{\partial s\partial t}\right|_{0}
					U[d+s\eta+t\zeta] \\
					=
					\left.\frac{\partial^2}{\partial t\partial s}\right|_{0}
					U[d+s\eta+t\zeta]
					=
					\delta^2U[d;\zeta,\eta].
				\end{multline}
				
				Now use the defining relation (\ref{HUd}):
				
				\begin{equation}
					\delta^2U[d;\eta,\zeta] = \int \eta_k(\mathcal H_U[d]\zeta)_k,~~
					\delta^2U[d;\zeta,\eta]=\int \zeta_k(\mathcal H_U[d]\eta)_k.
				\end{equation}
				
				Since the left sides are equal by (\ref{D2U}), the right sides are equal:
				
				\begin{equation}
					\int_\Omega \eta_k(\mathcal H_U[d]\zeta)_k\,{d^2x}
					=\int_\Omega \zeta_k(\mathcal H_U[d]\eta)_k\,{d^2x},
				\end{equation}
				
				which is exactly (\ref{SA}).
				
				\begin{equation}
					\Rightarrow\ \mathcal H_U[d]\ \text{is self-adjoint.}
				\end{equation}

				\subsection{Linearization and Eigenvalue Problem}
				
				In the linear stability analysis, our primary aim is to outline: ``Which deformation pattern wants to appear first?" Therefore, it is a selection problem. Selection is always governed by linear stability. Nonlinear terms compete after instability starts, but which mode appears first is determined by linear terms. 
				
				\subsection{Linearization at $d=0$}
				
				At $d=0$: $d_{k,\beta}=0$, $\sigma_{\alpha\beta}(0)=0$, and $M^{k\zeta}(0)=L^{k\zeta}$. Also
				
				\begin{equation}
					\delta u_{\gamma\delta}[0;\zeta]=\tfrac12(\zeta_{\gamma,\delta}+\zeta_{\delta,\gamma}). \nonumber
				\end{equation}
				
				Thus
				
				\begin{equation}
					(\mathcal H_U[0]\zeta)_k =
					-(\Sigma^{\alpha\beta}\zeta_{k,\beta})_{,\alpha}
					-(\delta\sigma_{\alpha k}[0;\zeta])_{,\alpha}
					+ L^{k\zeta}\zeta_\zeta,
				\end{equation}
				
				where
				
				\begin{equation}
					\delta\sigma_{\alpha k}[0;\zeta]=
					\tilde A_{\alpha k\gamma\delta}\tfrac12(\zeta_{\gamma,\delta}+\zeta_{\delta,\gamma}).
				\end{equation}

				\subsection{The instability operator}
				
				The equilibrium condition is,
				
				\begin{equation}
					E_k[d] = -\frac{\delta U}{\delta d}.
				\end{equation}
				
				Therefore its linearization is
				
				\begin{equation}\label{DEk}
					\delta E_k[0] = -\mathcal H_U[0].
				\end{equation}
				
				This is the operator used to study linear instability of the equilibrium equation (the ``soft mode'' of the PDE). So define the instability (Jacobian) operator
				
				\begin{equation}
					\mathbb J[0] := \delta E_k[0] = -\mathcal H_U[0].
				\end{equation}
				
				Applying minus to (\ref{DEk}):
				
				\begin{equation}
					(\mathbb J[0]\zeta)_k=
					+\big(\Sigma^{\alpha\beta}\zeta_{k,\beta}\big)_{,\alpha}
					+\big(\delta\sigma_{\alpha k}[0;\zeta]\big)_{,\alpha}
					-\;L^{k\zeta}\zeta_\zeta.
				\end{equation}
				
				\subsection{Fourier modes and the eigenvalue problem}
				
				We note that in the linearized stability analysis, the equilibrium operator (the equilibrium PDE) and the Hessian have constant coefficients, so its eigenfunctions are plane waves ($e^{i q \cdot {\bf x}}$) labeled by a continuous wavevector $q$. Plane waves are used because they diagonalize the linear operator as shown below. They identify the direction of the instability in the function space.
				\begin{equation}
					\zeta_k({\bf x})=\hat\zeta_k\,e^{i\B q\cdot {\bf x}},
					\qquad q_\alpha\in\mathbb R^2.
				\end{equation}
				Then $\partial_\alpha \mapsto i q_\alpha$. Compute each term carefully:
				
				(a) Prestress-gradient term
				\begin{equation}
					\big(\Sigma^{\alpha\beta}\zeta_{k,\beta}\big)_{,\alpha} \mapsto (i q_\alpha)\Sigma^{\alpha\beta}(i q_\beta)\hat\zeta_k
					=-(\Sigma^{\alpha\beta}q_\alpha q_\beta)\hat\zeta_k.
				\end{equation}
				(b) Elastic term
				
				First
				\begin{equation}
					\varepsilon_{\gamma\delta}(\zeta)=\tfrac12(iq_\delta \hat\zeta_\gamma + iq_\gamma \hat\zeta_\delta)e^{i\B q\cdot {\bf x}}.
				\end{equation}
				Then
				\begin{equation}
					\delta\sigma_{\alpha k}	=\tilde A_{\alpha k\gamma\delta}\varepsilon_{\gamma\delta}(\zeta)
					=\tilde A_{\alpha k\gamma\delta}\tfrac12(iq_\delta \hat\zeta_\gamma + iq_\gamma \hat\zeta_\delta)e^{i\B q\cdot {\bf x}}.
				\end{equation}
				Now divergence:
				\begin{equation}
					(\delta\sigma_{\alpha k})_{,\alpha}
					\mapsto
					(i q_\alpha)\delta\sigma_{\alpha k}
					=- \tilde A_{\alpha k\gamma\delta} ~q_\alpha q_\delta ~\hat\zeta_\gamma
				\end{equation}
				(using the minor symmetry $\tilde A_{\alpha k\gamma\delta}=\tilde A_{\alpha k\delta\gamma}$ to combine the two halves).
				
				(c) Local term
				\begin{equation}
					-L^{k\zeta}\zeta_\zeta \mapsto -L^{k\gamma}\hat\zeta_\gamma.
				\end{equation}
				\subsubsection{The unprojected Fourier eigenvalue matrix}
				
				So $\mathbb J[0]\zeta=\lambda \zeta$ becomes
				\begin{equation}\label{EP}
					\Big[(\Sigma^{\alpha\beta}q_\alpha q_\beta)\,\delta_{k\gamma}+\tilde A_{\alpha k\gamma\delta}\,q_\alpha q_\delta- L^{k\gamma}
					\Big]\hat\zeta_\gamma=\lambda(q)\,\hat\zeta_k.
				\end{equation}
				This is the exact, unprojected Jacobian eigenvalue problem. Instability occurs when $ \lambda(q)$ first reaches zero at some critical wavevector $q_0$.
				
				Let us define
				
				\begin{equation}
					\B q=|q|\B m,\qquad \B m=(\cos\phi \B n,\sin\phi  \B t) \equiv (m_\alpha, m_\beta) \ ,
				\end{equation}
				with $\B e\cdot \B m=0$.
				
				\subsubsection{The prestress projection carrying the $(\theta)$-dependence}
				
				We defined the scalar,
				\begin{equation}
					\Sigma^{\alpha\beta}(m):=\Sigma^{\alpha\beta}m_\alpha m_\beta
				\end{equation}
				so that,
				\begin{equation}\label{id}
					\Sigma^{\alpha\beta}q_\alpha q_\beta =\Sigma^{\alpha\beta}m_\alpha m_\beta \,|q|^2 =\Sigma(m)  \,|q|^2. 
				\end{equation}
				
				The elastic part in (\ref{EP}) is
				$\tilde A_{\alpha k\zeta\delta}q_\alpha q_\delta
				$. Insert this into (\ref{elten}) with $q_\alpha=|q| n_\alpha$:
				
				\begin{multline}
					\tilde A_{\alpha k\zeta\delta}q_\alpha q_\delta=
					\lambda\,(\delta_{\alpha k}q_\alpha)(\delta_{\zeta\delta}q_\delta)	\\ +\mu\left(\delta_{\alpha\zeta}\delta_{k\delta}q_\alpha q_\delta+\delta_{\alpha\delta}\delta_{k\zeta}q_\alpha q_\delta\right).
				\end{multline}
				
				Now simplify term by term:
				
				\begin{itemize}
					\item 	 $\delta_{\alpha k}q_\alpha = q_k$	 
					\item	 $\delta_{\zeta\delta}q_\delta = q_\zeta$	 
					\item	 So the $\lambda$-term gives $\lambda\, q_k q_\zeta$.
				\end{itemize}

				For the $\mu$-terms:
				
				\begin{align}
					\delta_{\alpha\zeta}\delta_{k\delta}q_\alpha q_\delta &= q_\zeta q_k \nonumber \\
					\delta_{\alpha\delta}\delta_{k\zeta}q_\alpha q_\delta &= (q_\alpha q_\alpha)\delta_{k\zeta} = (\B q\cdot\B  q)\delta_{k\zeta}=|q|^2\delta_{k\zeta}
				\end{align}
				
				Therefore
				
				\begin{equation}
					\tilde A_{\alpha k\zeta\delta}q_\alpha q_\delta=
					\mu\,|q|^2\,\delta_{k\zeta} +(\lambda+\mu)\,q_k q_\zeta.
				\end{equation}
				
				Since $q_k=\kappa m_k$, we can write $q_k q_\zeta = \kappa^2 m_k m_\zeta$:
				
				\begin{equation}\label{Abar}
					\tilde A_{\alpha k\zeta\delta}q_\alpha q_\delta=|q|^2\Big[\mu\,\delta_{k\zeta}+(\lambda+\mu)\,m_k m_\zeta\Big].
				\end{equation}
				
				\subsection{Computing the full $2\times2$ eigenvalue matrix}
				
				We consider the isotropic systems such that the ``local'' tensor is isotropic:
				
				\begin{equation}\label{LA}
					L^{k\zeta}=\C A\,\delta^{k\zeta}.
				\end{equation}
				
				Now combine (\ref{id}), (\ref{Abar}), (\ref{LA}) into (\ref{EP}):
				
				\begin{equation}
					\mathsf M(\kappa,m)
					=\kappa^2\Big[\Sigma(m)\,I +\mu\,I +(\lambda+\mu) \B m\otimes\B m \Big] -\C A\,I. 
				\end{equation}
				Here, $\B m\otimes \B m$ is the dyadic product. It is a second-rank tensor with components, $(\B m\otimes \B m)_{ij} = m_i m_j$. It acts as a projection operator onto the direction $\B m$. When it is applied to a vector $\B v$, it projects the component of $\B v$ along $\B m$, i.e., $(\B m\otimes \B m){\B v} = (\B m\cdot \B v)\B m$. So the eigen-problem is
				
				\begin{equation}
					\mathsf M(\kappa,m)\,\hat\zeta = \lambda(\kappa,m)\,\hat\zeta.
				\end{equation}
				Note that the matrix is diagonal in the basis
				of the longitudinal direction $\B m$ and the transverse direction $\B e$.
				Use $(\B m\otimes\B  m)\B e =\B m (\B m\cdot \B e)=0$. So a general $\hat\zeta$ can be decomposed in these two directions, leading to 
				
				\begin{equation}
					\mathsf M \B e = \Big[|q|^2(\Sigma(m) +\mu)-\C A\Big]\B e\ , 
				\end{equation}
				and 
				\begin{equation}
					\lambda_T(|q|,m)=\big(\Sigma(m)+\mu\big)|q|^2 - \C A,\qquad \hat\zeta = \B e.
				\end{equation}
				Similarly
				\begin{equation}
					\mathsf M \B m = \Big[|q|^2(\Sigma(m)+\mu+(\lambda+\mu))-\C A\Big]\B m.
				\end{equation}
				Therefore
				\begin{equation}
					\lambda_L|q|,m)=\big(\Sigma(m)+\lambda+2\mu\big)|q|^2 - \C A,\qquad \hat\zeta = \B m.
				\end{equation}
				
				Thus, we find that, in the unprojected linearized Hessian, each wave vector direction has two eigenmodes corresponding to the longitudinal mode ($\lambda_L$), and the transverse mode with eigenvalue ($\lambda_T$). These eigenvalues correspond to the rigidity of the material against deformation with respect to longitudinal and transverse displacements, respectively. 
				The longitudinal component contains the bulk contribution ($\lambda+2\mu$), whereas the transverse component contains only the shear modulus ($\mu$). Since ($\lambda+2\mu>\mu$) for any elastic solid, we always have ($\lambda_T<\lambda_L$). Now, as the load increases, the transverse component $\lambda_T$ goes to zero first and becomes the soft-mode. Since $\lambda_T$ only represents shear resistance, physically it corresponds to the shear deformation. Therefore, the $\lambda_T$ represents the shear polarization.

				\subsubsection{``Shear polarization'' as the transverse direction}
				
				Here we can understand how ``polarization'' decides the nature of eigenvalue problem. In the unprojected linearized Fourier eigenproblem, we assume a plane-wave perturbation
				
				\begin{equation}
					\zeta_k({\bf x})=\hat\zeta_k e^{i \B q\cdot {\bf x}}.
				\end{equation}
				
				\(q\) sets the direction along which the field varies (the wavevector).  
				\(\hat\zeta\) is a constant vector that sets the direction of displacement of that mode.  
				That direction \(\hat\zeta\) is what we call the \emph{polarization}.
				
				\subsubsection{Strain produced by a plane wave:}
				
				For simplicity let us consider small strain of the perturbation which is
				\begin{equation}
					u_{\alpha\beta}(\zeta)=\frac12(\zeta_{\alpha,\beta}+\zeta_{\beta,\alpha}).
				\end{equation}
				For $\zeta_\alpha=\hat\zeta_\alpha e^{iq\cdot {\bf x}}$,
				$
				\zeta_{\alpha,\beta}= i q_\beta \hat\zeta_\alpha e^{i\B q\cdot {\bf x}},
				$
				so
				\begin{equation}
					u_{\alpha\beta}=\frac{i}{2}\Big(q_\beta \hat\zeta_\alpha+q_\alpha \hat\zeta_\beta\Big)e^{iq\cdot {\bf x}}. \label{A}
				\end{equation}
				
				Now decompose $\hat\zeta$ into components parallel and perpendicular to $\B q$,
				$ \hat\zeta = a\,\B m + b\,\B e$.
				
				\subsubsection{Transverse choice gives pure shear:}
				
				Let us take transverse polarization: $\hat\zeta=b \B e$ and compute the normal strain in the propagation direction $m$, denoted as $u^{(mm)}$:
				\begin{equation}
					u^{(mm)}=m_\alpha m_\beta u_{\alpha\beta}.
				\end{equation}
				Using Eq.~(\ref{A}):
				\begin{equation}
					u^{(mm)}=\frac{i}{2}m_\alpha m_\beta (q_\beta \hat\zeta_\alpha+q_\alpha \hat\zeta_\beta)e^{iq\cdot {\bf x}}.
				\end{equation}
				Insert $q_\beta=|q| m_\beta$:
				\begin{align}
					u^{(mm)} &=\frac{i|q|}{2}m_\alpha m_\beta (m_\beta \hat\zeta_\alpha+m_\alpha \hat\zeta_\beta)e^{i\B q\cdot x} \nonumber \\
					&=\frac{i|q|}{2}\Big (\B m\cdot\hat\zeta)+(\B m\cdot\hat\zeta)\Big]e^{i\B q\cdot {\bf x}}.
				\end{align}
				So
				\begin{equation}
					u^{(mm)}= i|q| (\B m\cdot\hat\zeta)e^{i\B q\cdot {\bf x}}.
				\end{equation}
				But transverse means $m\cdot\hat\zeta=0$. Therefore
				\begin{equation}
					u^{(mm)}=0 \quad\text{for transverse polarization.} \label{B}
				\end{equation}
				Similarly, the normal strain along $\B e$,
				
				\begin{equation}
					u^{(ee)}:=e_\alpha e_\beta u_{\alpha\beta} =0,
				\end{equation}
				also vanishes for the same reason (because $\B e\cdot q=0$).
				
				But the shear component
				\begin{equation}
					u^{(me)}=m_\alpha e_\beta u_{\alpha\beta}
				\end{equation}
				does not vanish; using (\ref{A}) and $q=|q| m$, $\hat\zeta=b \B e$, we get
				
				\begin{equation}
					u^{(me)}=u^{(em)}=\frac{i|q| b}{2}e^{i\B q\cdot {\bf x}}\neq 0. \label{D}
				\end{equation}
				
				Therefore transverse polarization produces strain with only off-diagonal components in the $(\B m,\B e)$ basis. This is pure shear.
				
				\subsubsection{The longitudinal choice is not shear}
				
				If $\hat\zeta=a \B m$ (longitudinal), then $\B m\cdot\hat\zeta=a\neq 0$ and from the same calculation:
				\begin{equation}
					u^{(mm)}= i|q| a\,e^{i\B q\cdot {\bf x}}\neq 0,
				\end{equation}
				i.e. it produces compression or extension along $\B m$. That is not shear. So, kinematically
				transverse polarization($\hat\zeta\perp q)$ means shear strain.
				
				A plane-wave mode has two independent directions; the wavevector $\B q$, tells us the direction along which the field varies, and the displacement amplitude $\hat\zeta$, which tells the direction the material points move. The strain is the symmetric gradient of displacement. If $\hat\zeta\perp \B q$ (transverse), all normal components vanish in the \((\B m,\B e)\) basis and only the mixed component $u^{(me)}$ remains; that is exactly pure shear.

				\subsection{Applied Macroscopic Shear and the Shear Polarization Selected by Hessian}
				
				To avoid any confusion between the ``applied shear direction'' and 
				the ``shear polarization selected by the Hessian'' here we outline the basic difference between 
				these two quantities.
				
				The applied shear is a macroscopic quantity. When we apply a pure shear stress
				$
				\Sigma^{\alpha\beta} =\Sigma^{12} = \Sigma 
				$.
				This fixes the principal stress directions, loading axes, etc. This is a boundary/loading condition. Therefore, this shear is given.
				On the other hand, the shear polarization of a Hessian mode is an outcome of the eigenproblem of the unprojected Hessian. Therefore, this is not an assumed or imposed shear. This shear emerges as an eigenvector.\\
				
				From the previous discussion we have the unprojected linearized eigenproblem:
				\begin{equation}
					\mathsf M(\kappa,m)~\hat d = \lambda(\kappa,m)~\hat d
				\end{equation}
				Here, $\hat d$ is a vector displacement amplitude, whereas $(\lambda)$ is a stiffness/rigidity representing the eigenvalue, and $m$ fixes the wavevector direction $\B m$. Once we have this, the eigenvalue problem of the Hessian corresponding to the deformation is: ``If the system tries to deform locally with wavevector $\B m$, in which direction does it want to move?'' That direction is the polarization. In 2D, for a given wavevector direction $\B m$, there are exactly two independent displacement polarizations: 
				
				\begin{itemize}
					\item Longitudinal polarization $\hat d \parallel \B m$ $\rightarrow$ compression/dilation,
					\item Transverse polarization $\hat d \perp \B m  \rightarrow$ shear deformation.
				\end{itemize}
				These are  eigenvectors of the Hessian. 
				
				Thus, the shear polarization is selected by the Hessian implies that among all possible infinitesimal displacement fluctuations, the one whose displacement vector is perpendicular to the wavevector is the softest eigenmode of the Hessian.
				
				Therefore, we do not assume shear localization beforehand. We apply a macroscopic shear stress. The Hessian then shows us that the softest fluctuation is transversely polarized, implying that displacement is perpendicular to the band normal. Strain is a shear-like localization that produces a shear band.

				\subsection{Reduction of Displacement Field Along the Soft-Mode}
				
				Note that, as shown earlier, the linearized Hessian $\mathcal{H}_{\rm lin}$ is a self-adjoint operator acting on vector fields. It has an eigenbasis
				\begin{equation}
					\mathcal{H}_{\rm lin}\B \psi^{(n)}=\lambda_n \B \psi^{(n)},
					\qquad
					\langle \B  \psi^{(m)},\B \psi^{(n)}\rangle=\delta_{mn},
				\end{equation}
				where
				$
				\langle \B a,\B b\rangle:=\int_\Omega a_k({\bf x})b_k({\bf x})\,{d^2x}.
				$ Then any admissible displacement can be expanded as:
				
				\begin{equation}
					d_k({\bf x})=\sum_{n} a_n\,\psi^{(n)}_k({\bf x}), \qquad a_n=\langle \B \psi^{(n)},\B d\rangle.
					\label{basis}
				\end{equation}
				
				This is the mathematical meaning of ``displacement has a component along a Hessian mode.'' Now we outline the meaning of each symbol.

				\subsubsection{Computing the quadratic energy in eigenmode coordinates}
				
				We want to compute:
				\begin{equation}
					\langle\B  d,\mathcal{H}_{\rm lin} \B d\rangle.
				\end{equation}
				To compute this we use the expansion for $\B d$, 
				\begin{equation}
					\B d=\sum_n a_n\B \psi^{(n)}, \qquad \mathcal{H}_{\rm lin} \B d = \sum_n a_n \mathcal{H}_{\rm lin} \B \psi^{n} = \sum_n a_n\lambda_n\B \psi^{n}.
				\end{equation}
				Taking inner product
				\begin{equation}
					\langle \B d,\mathcal{H}_{\rm lin} \B d\rangle = \left\langle \sum_m a_m\psi^m,\ \sum_n a_n\lambda_n\psi_n\right\rangle
					= \sum_{n\ge1}\lambda_n a_n^2 \ . 
				\end{equation}
				Note that in above equation each term $\lambda_n a_n^2$ is the quadratic energy contribution of mode $n$. Large $\lambda_n$ $\rightarrow$ costly to excite, small $\lambda_n$ $\rightarrow$ cheap to excite, and $\lambda_n = 0$ $\rightarrow$ costs no quadratic energy.
				
				Thus an instability is signalled by 
				\begin{equation}
					\lambda_1 \approx 0 \quad\text{at the threshold} \ . 
				\end{equation}
				We will assume below that the next mode stays positive:
				\begin{equation}
					\lambda_2 \ge \lambda_D> 0. 
				\end{equation}
				This is the spectral gap assumption.
				\subsubsection{Inequality for the soft-mode dominance}
				We start from the exact identity:
				\begin{equation}\label{IN1}
					\langle \B d,\mathcal{H}_{\rm lin} \B d\rangle= \lambda_1 a_1^2 + \sum_{n\ge2}\lambda_n a_n^2.  
				\end{equation}
				
				Now use the spectral difference: for every $n\ge2$, $\lambda_n\ge\lambda_{\rm D}$. Therefore:
				
				\begin{equation}\label{IN2}
					\sum_{n\ge2}\lambda_n a_n^2\ge\sum_{n\ge2}\lambda_{\rm D}\,a_n^2
					= \lambda_{\rm D}\sum_{n\ge2}a_n^2.
				\end{equation}
				
				Insert (\ref{IN2}) into (\ref{IN1}):
				
				\begin{equation}\label{IN3}
					\langle \B d,\mathcal{H}_{\rm lin} \B d\rangle \ge \lambda_1 a_1^2 + \lambda_{\rm D}\sum_{n\ge2}a_n^2. 
				\end{equation}
				
				Now isolate the higher-mode contribution. Subtract $\lambda_1 a_1^2$:
				
				\begin{equation}\label{IN4}
					\langle \B d,\mathcal{H}_{\rm lin} \B d\rangle - \lambda_1 a_1^2 \ge \lambda_{\rm D}\sum_{n\ge2}a_n^2.
				\end{equation}
				
				Divide by $\lambda_{\rm D}$:
				
				\begin{equation}\label{IN5}
					\sum_{n\ge2}a_n^2\le\frac{1}{\lambda_{\rm D}}
					\big(\langle \B d,\mathcal{H}_{\rm lin} \B d\rangle - \lambda_1 a_1^2\big). 
				\end{equation}

				This is the inequality representing the constraint on the modes other than the soft one. We note that the term $\sum_{n\ge2}a_n^2$ is the squared norm of the ``stiff'' modes. We can represent it by $\B d^{\rm s}$, where,
				
				\begin{equation}
					\B d^{\rm s}:= \sum_{n\ge2}a_n\B \psi_n,
					\qquad
					(d^{\rm s})^2 = \sum_{n\ge2}a_n^2.
				\end{equation}
				
				The quantity $ (d^{\rm s})^2$ measures how much of the displacement lives outside the soft eigen-space. On the other hand, $\langle \B d,\mathcal{H}_{\rm lin} \B d\rangle$ is the quadratic energy. It measures the amount of quadratic energy the displacement costs.
				Thus, the inequality tells us that if the quadratic energy is small, then the amplitude outside the soft mode must be small. This is the precise mathematical statement of ``soft-mode dominance''. In other words, this means that  
				\begin{equation}\label{SMD}
					d_k({\bf x}) \approx a_1 \psi_k^{(1)}({\bf x}). 
				\end{equation}
				We note that the spectral difference $\lambda_{\rm D}$ ensures that only one displacement pattern is soft at the instability; all others remain costly. Since all stiff modes cost finite quadratic energy, any low-energy displacement must suppress them, leaving only the soft mode.

				\subsection{Structure of the soft eigenmode $\B \psi^{(1)}_k({\bf x})$}
				The linear Hessian $\mathcal{H}_{\rm lin}$ has constant coefficients since we still have $d=0$.
				Therefore the eigenmodes can be chosen as plane waves
				\begin{equation}
					\psi_{k}^{(1)} ({\bf x})= \hat{\psi}_k ~e^{i\B q\cdot {\bf x}}
				\end{equation}
				Since $\B q = |q| \B n$, $\B q\cdot {\bf x} = |q| (\B m \cdot {\bf x})$.
				
				We define a new scalar coordinate $\xi$, such that $\B m \cdot {\bf x} = \xi$.
				Therefore the soft-mode is simplified to
				\begin{equation}
					\B \psi_{k}^{(1)} ({\bf x})= \hat{\psi}_k ~e^{i |q| \xi}.
				\end{equation}
				At the instability, $\lambda_1 ~ \rightarrow 0$, the soft-mode represents a ``soft-subspace''; this does not select a single function in space but identifies a direction $\B m$ and a polarization $\B e$ of the soft eigen-space. At the instability, the linear operator becomes singular and fails to fix the spatial profile uniquely, so linear theory alone is insufficient. When nonlinear terms are retained, superpositions of plane waves with the same direction $\B m$ but closeby wave-numbers $|q|$ are naturally allowed, and these combine into a slowly varying amplitude profile $\phi(\xi)$ with $\xi = \B m \cdot {\bf x}$. 
				Thus, at post-instability or near instability, we obtain a localized envelope. Thus, at the instability, the general behaviour of the soft-mode is described by the following form
				\begin{equation}
					\psi^{(1)}_k({\bf x}) = e_k\,\phi(\xi),
				\end{equation}
				where, $e_k$= constant polarization vector. $\xi$ = coordinate along some direction, $\phi(\xi)$ = scalar function whose form will be selected by the 
				nonlinear saturation or after solving the projected nonlinear equilibrium equation. Thus, the plane waves are not the shear band itself but the basis used to identify the soft mode. So it is important to note that the emergence of the shear band is the nonlinear continuation of the linear soft-mode localized across the normal direction $\B m$, not the linear mode itself.\\
				\subsubsection*{substituting the soft-mode form into the truncated expansion:}
				To get the final structure of the displacement field leading to the shear band profile, we substitute the structure of the soft-mode derived above into the following equation for the displacement field along the sub-space defined by the zero-mode, which we have from (\ref{SMD}),
				\begin{equation}
					d_k({\bf x}) \approx a_1\,\psi^{(1)}_k({\bf x})
				\end{equation}
				Using:
				\begin{equation}
					\psi^{(1)}_k({\bf x}) = e_k\,\phi(\xi)
				\end{equation}
				Define:
				\begin{equation}
					f(\xi) = a_1\,\phi(\xi)
				\end{equation}
				Then:
				\begin{equation}\label{Ansatz}
					d_k({\bf x}) = e_k\,f(\xi) 
				\end{equation}
				This has two parts: vector part $e_k$: fixed polarization, spatial part $\phi(\xi)$: scalar function. This is exactly our starting equation for a scalar field. The vector nature is captured by $e_k$ with a single scalar field. Eq.~ (\ref{basis}) says ``any displacement is a sum of eigen-modes''. Near the instability, only the soft mode matters, and its vector structure factorizes into a fixed polarization and a scalar amplitude profile, giving our starting point. 
				
				It is important to note that all vector physics is still present in $e_k$. The unprojected Hessian eigenproblem determines the direction and nature of deformation. The scalar equation only describes: how the soft mode localizes. So, Eq. (\ref{Ansatz}) is not an assumption about shear, it is rather the standard form of the vector instability.
				
				Note that the superscript ($1$) in $d_k({\bf x}) = a \psi^{(1)}_k({\bf x})$ represents the zero-mode or the soft-mode corresponding to the zero eigenvalue of the linearized unprojected Hessian. However, it does not represent a unique eigenfunction. The zero mode, at the instability, is degenerate and consists of all the displacement fields that are of the form written as $d_k({\bf x}) = e_k f(\xi)$, where $e_k$ represents the polarization fixed by the eigenmode, and $\xi=\B m \cdot {\bf x}$. The $f(\xi)$ is an arbitray scalar function at this stage. Therefore,  $d_k({\bf x}) = e_k f(\xi)$, in a short form, stands for projecting the full displacement field into this soft subspace and neglecting the stiff or non-zero modes that cost energy. We note that the actual spatial form of the function $f(\xi)$, where $\xi =\B m \cdot {\bf x}$, is not determined by the linear theory and is determined only after incorporating the nonlinear effects.
				\section{RHS of Eq.(\ref{finalNL})}
				\label{RHS}
				
				The RHS of Eq.~(\ref{finalNL}) reads
				
				\begin{equation}
					RHS_k = -L_{k\zeta}\,d_\zeta \;+\; T_{kpqr}\,d_p d_q d_r .
				\end{equation}
				
				with:
				
				\begin{equation}
					L_{k\zeta}=\Gamma^k_{\alpha} \Lambda_{\alpha\beta}\Gamma^\zeta_{\beta},
				\end{equation}
				
				\begin{equation}
					T_{kpqr}=\Gamma^k_{\alpha} \Lambda_{\alpha\beta}\,G_{\beta\gamma\delta\eta}\,\Lambda_{\gamma\mu}\Gamma^r_{\mu}\,
					\Lambda_{\delta\nu}\Gamma^p_{\nu}\,
					\Lambda_{\eta\tau}\Gamma^q_{\tau}.
					\label{T-def}
				\end{equation}

				\subsection{Projection along a soft-mode}
				
				We have 
				
				\begin{equation}
					d_k(\mathbf{x}) = e_k\,f(\xi),
					\qquad \text{with } \xi=\B m\cdot x \ .
					\tag{ansatz}
				\end{equation}
				Then, we have
				\begin{equation}
					d_\zeta = e_\zeta f(\xi),
					\qquad
					d_p d_q d_r = e_p e_q e_r\,f^3(\xi).
					\tag{products}
				\end{equation}
				Substitute this into RHS$_k$ and separate powers of $u$,
				\begin{equation}
					-L_{k\zeta}d_\zeta = -L_{k\zeta}(e_\zeta f(\xi))=-(L_{k\zeta}e_\zeta)\,f(\xi) \ ,
					\tag{linear}
				\end{equation}
				\begin{equation}
					T_{kpqr}d_p d_q d_r
					=(T_{kpqr}e_p e_q e_r)\,f^3(\xi) \ .
					\tag{cubic}
				\end{equation}
				
				So altogether,
				
				\begin{equation}
					\text{RHS}_k
					= -(L_{k\zeta}e_\zeta)\,f(\xi)
					\;+\;
					(T_{kpqr}e_p e_q e_r)\,f^3(\xi).
				\end{equation}
				Now project along the same mode, multiply by $e_k$.
				This defines the scalar RHS:
				
				\begin{equation}
					e_k\,\text{RHS}_k
					= -\,e_k(L_{k\zeta}e_\zeta)\,f(\xi)
					\;+\;
					e_k(T_{kpqr}e_p e_q e_r)\,f^3(\xi) \ .
				\end{equation}
				This the definitions of the parameters $\C A$ and $\C B$ are
				\begin{equation}
					\C A \equiv e_k L_{k\zeta} e_\zeta,\qquad \C B\equiv e_k T_{kpqr} e_p e_q e_r.
				\end{equation}
				Substituting $L$ and $T$ from Eqs.~(\ref{QP26}) we get the explicit tensor contractions
				\begin{equation}
					\C A= e_k\,\Gamma^k_{\alpha} \Lambda_{\alpha\beta}\Gamma^\zeta_{\beta}\,e_\zeta \ ,
				\end{equation}
				\begin{equation}
					\C 	B =
					e_k\,
					\Gamma^k_{\alpha} \Lambda_{\alpha\beta}\,
					G_{\beta\gamma\delta\eta}\,
					\Lambda_{\gamma\mu}\Gamma^r_{\mu}\,
					\Lambda_{\delta\nu}\Gamma^p_{\nu}\,
					\Lambda_{\eta\tau}\Gamma^q_{\tau}\;
					e_p e_q e_r \ .
				\end{equation}
				
				\subsection{Isotropic and homogenous systems}
				In isotropic and homogeneous systems we can use the following forms
				\begin{equation}
					\Gamma^k_{\alpha}=g\delta^k_{\alpha},\qquad
					\Lambda_{\alpha\beta}=\Lambda\delta_{\alpha\beta},
					\qquad
					\Lambda^{\alpha\beta}=\frac{1}{\Lambda}\delta^{\alpha\beta} \ .
				\end{equation}
				Regarding the isotropic form of the tensor $G_{\alpha\beta\gamma\delta}$ one could think that it involves two parameters like the elastic tensor Eq.~(\ref{elten}). In fact this tensor is fully symmetric, since it originates from a fourth order derivative, 	
				\begin{equation}
					G_{\alpha\beta\gamma\delta}=\frac{\partial^4 W_4}{\partial P_\alpha\partial P_\beta\partial P_\gamma\partial P_\delta} \ , 
				\end{equation}
				where
				\begin{equation}
					W_4(P)=\frac{h}{4}(P_\alpha P_\alpha)^2\ .
				\end{equation}
				Therefore the tensor is fully symmetric and can be written as
				\begin{equation}
					G_{\alpha\beta\gamma\delta}
					= h\left(\delta_{\alpha\beta}\delta_{\gamma\delta}
					+ \delta_{\alpha\gamma}\delta_{\beta\delta}
					+ \delta_{\alpha\delta}\delta_{\beta\gamma}\right) \ .
					\label{Hsym}
				\end{equation}
				Thus in isotropic and homogeneous systems the estimate of the parameters $\C A$ and $\C B$ simplifies further.
				\subsubsection{Computing $\C A$}
				wrting again
				\begin{equation}
					L_{k\zeta}=\Gamma^k_{\alpha}\Lambda^{\alpha\beta}\Gamma^\zeta_{\beta}.
				\end{equation}
				we substitute the isotropic forms
				\begin{equation}
					L_{k\zeta} = (g\delta^k_{\alpha})\left(\frac{1}{\Lambda}\delta^{\alpha\beta}\right)(g\delta^\zeta_{\beta}) = \frac{g^2}{\Lambda}\delta^{k\zeta} \ ,
				\end{equation}
				\begin{equation}
					\C	A=e_k L_{k\zeta} e_\zeta
					= \frac{g^2}{\Lambda} e_k \delta^{k\zeta} e_\zeta
					= \frac{g^2}{\Lambda}.
				\end{equation}
				Thus, 
				\begin{equation}
					\boxed{\C A = g^2/\Lambda }
				\end{equation}
				
				At this point we recall that the tensor $\B L$ appeared already in Eqs.(\ref{L1}) and (\ref{finalNL}).
				In previous {\em linear} work it was identified in terms of the screening lengths, reading
			\begin{align} 
				\B {\widetilde K} =
				\begin{bmatrix}
					\bf    \tilde{\kappa}_{e}^{2} & \bf-\tilde{\kappa}_{o}^{2}  \\
					\bf    \tilde{\kappa}_{o}^{2} &\bf \tilde{\kappa}_{e}^{2} 
				\end{bmatrix},
			\end{align}
			which can be written in its symmetric and antisymmetric parts 
			
			\begin{equation}
				\widetilde K_{\alpha\beta} = \tilde{\kappa}_e^2 \delta^{\alpha\beta} + \tilde{\kappa}_o^2 \begin{pmatrix} 0 & 1 \\ -1 & 0 \end{pmatrix} 
				= \tilde{\kappa}_e^2 \delta^{\alpha\beta} + \tilde{\kappa}_o^2\epsilon^{\alpha\beta}
			\end{equation}
			where, $\delta^{\alpha\beta}$ is tensor of Kronecker delta and $\epsilon_{\alpha\beta}$ is Levi civita tensor with ($\epsilon_{21} = - \epsilon_{12}$).
			Now, with the above form of $\B {\widetilde K}$, if it continues to be valid in the nonlinear environment,  we compute the projection,
			\begin{align}
				\C	A &=e_\alpha L_{\alpha\beta} e_\beta \nonumber \\
				&=  \tilde{\kappa}_e^2 e_\alpha \delta^{\alpha\beta} e_\beta + \tilde{\kappa}_o^2 e_\alpha \epsilon^{\alpha\beta} e_\beta 
				= \tilde{\kappa}_e^2(e\cdot e).
			\end{align}
			For the normalized mode ($e\cdot e=1$), with the equation (\ref{SM}), we obtain:
			
			\begin{equation}
				\boxed{\C A = \tilde{\kappa}_e^2},~~ \text{and}~~ \kappa_e^2 =\mu \tilde{\kappa}_e^2 .
			\end{equation}

				\subsubsection{Compute $B$}
				We start with
				\begin{equation}
					T_{kpqr}=\Gamma^k_{\alpha}\Lambda^{\alpha\beta}
					G_{\beta\gamma\delta\eta}\Lambda^{\gamma\mu}\Gamma^{r}_{\mu}\Lambda^{\delta\nu}\Gamma^{p}_{\nu}\Lambda^{\eta\tau}\Gamma^{q}_{\tau}.
				\end{equation}
				Insert $\Gamma^k_{\alpha}=g\delta^k_{\alpha}$ and $\Lambda^{\alpha\beta}=\frac{1}{\Lambda}\delta^{\alpha\beta}$ everywhere.
				Then
				\begin{equation}
					T_{kpqr}
					= \frac{g^4}{\Lambda^4}
					\left(\delta^k_{\alpha}\delta^{\alpha\beta}\right)
					G_{\beta\gamma\delta\eta}
					\left(\delta^{\gamma\mu}\delta^r_{\mu}\right)
					\left(\delta^{\delta\nu}\delta^p_{\nu}\right)
					\left(\delta^{\eta\tau}\delta^q_{\tau}\right).
				\end{equation}
				Each parenthesis now reduces to:
				\begin{equation}
					\delta^k_{\alpha}\delta^{\alpha\beta}=\delta^{k\beta},\quad
					\delta^{\gamma\mu}\delta^r_{\mu}=\delta^{\gamma r},\quad
					\delta^{\delta\nu}\delta^p_{\nu}=\delta^{\delta p},\quad
					\delta^{\eta\tau}\delta^q_{\tau}=\delta^{\eta q}.
				\end{equation}
				Thus
				\begin{equation}
					T_{kpqr}= \frac{g^4}{\Lambda^4} G_{k r p q}.
					\tag{*}
				\end{equation}
				Now by definition,
				\begin{equation}
					\C	B=e_k T_{kpqr} e_p e_q e_r = \frac{g^4}{\Lambda^4} e_k G_{k r p q} e_r e_p e_q.
				\end{equation}
				Now use the isotropic $\B G$ Eq.~(\ref{Hsym}).
				Performing all the contraction we finally find
				\begin{equation}
					\boxed{\C B=\frac{3h g^4}{\Lambda^4}.}
				\end{equation}

				\section{Hessian is self adjoint}
				\label{self}
				
				Here we show that the Hessian operator $\mathcal{H}$ derived from the second variation of the energy functional is self-adjoint. This ensures real eigenvalues, which correspond to normal mode frequencies or stability exponents. Self-adjointness guarantees that the system’s energy is real and that time evolution (if considered) preserves norm. Self-adjoint operators have a complete set of orthogonal eigenfunctions, which allows us to expand arbitrary perturbations in eigenmodes, identify instabilities through the sign of the smallest eigenvalue, and use variational methods to estimate critical stresses.

				\subsubsection*{Start from the second variation in the paper}
				
				From the explicit Taylor expansion in Equation (71) in the draft the second variation about an equilibrium solution $ f(\xi) $ is
				
				\begin{equation}\label{SA1}
					\delta^2F[f;\eta] = \int\left[ a(\xi)(\eta')^2 + V(\xi)\eta^2 \right] d\xi,
				\end{equation}
				
				where, $\eta =\delta f$, and $\eta^{\prime} = \delta f^{\prime}$, and 
				
				\begin{eqnarray}\label{SA2}
					a(\xi)&=& \mu + \Sigma(m) + \frac{3}{2}(\lambda+2\mu)(f'(\xi))^2 \ ,\nonumber\\
					V(\xi) &=& -\C A + 3\C B f^2(\xi) \ .
				\end{eqnarray}
				Here $a(\xi)$ is the strain-dependent stiffness (since it depends on $f'(\xi)$).
				\subsubsection{Convert the quadratic form into an operator $H$}
				
				The general rule is: an operator $H$ is associated to $\delta^2F $ if
				\begin{equation}\label{SA3}
					\delta^2F[u;\eta] = \int \eta(\xi) (H\eta)(\xi) d\xi
				\end{equation}
				(up to a conventional factor $ 1/2 $). Start with the first term:
				\begin{equation}\label{SA4}
					\int a(\xi) (\eta')^2 d\xi.
				\end{equation}
				Integrating by parts once:
				\begin{equation}\label{SA5}
					\int a (\eta')^2 d\xi = \int a \eta' \eta' d\xi = \Big[a \eta \eta'\Big]_{\partial} - \int \eta \frac{d}{d\xi}\big(a \eta'\big) d\xi.
				\end{equation}
				Under periodic BCs or variations vanishing at boundaries, the boundary term is zero, so:
				\begin{equation}\label{SA6}
					\int a (\eta')^2 d\xi = -\int \eta \frac{d}{d\xi}\big(a \eta'\big) d\xi.
				\end{equation}
				Therefore (\ref{SA3}) becomes
				\begin{equation}\label{SA7}
					\delta^2F[f;\eta] = \int \eta\left[ -\frac{d}{d\xi}\left(a(\xi)\frac{d\eta}{d\xi}\right) + V(\xi)\eta \right] d\xi.
				\end{equation}
				So the Hessian operator is
				\begin{equation}\label{SA8}
					\mathcal{H} = -\frac{d}{d\xi}\left( a(\xi)\frac{d}{d\xi} \right) + V(\xi), 
				\end{equation}
				with $ a(\xi), V(\xi) $ given by (\ref{SA2}). This is exactly the self-adjoint Sturm–Liouville form under Dirichlet boundary conditions.”
				
				\subsubsection*{Self-adjointness with $a(\xi)$ dependence on strain}
				
				Self-adjointness is a statement about the operator acting on test functions $\eta$, not about whether coefficients depend on the background solution $f(\xi)$. Here $a(\xi)$ is a given real function of $\xi$ once we fix the equilibrium profile $f(\xi)$. It doesn’t matter that $a$ came from $u'$; for the linearized problem it is just a real coefficient function.\\
				
				\textit{Explicit self-adjointness check:} Let us take two smooth test functions $\phi(\xi)$ and $\psi(\xi)$ satisfying the same boundary conditions (vanishing at boundaries), and consider the inner product,
				\begin{equation}\label{SA9}
					\langle \phi,\psi\rangle = \int \phi(\xi)\psi(\xi) d\xi.
				\end{equation}
				Compute $ \langle \phi, H\psi\rangle $ using (\ref{SA8}):
				\begin{equation}\label{SA10}
					\langle \phi, \mathcal{H}\psi\rangle = \int \phi\left[ -\frac{d}{d\xi}\left(a \psi'\right) + V\psi \right] d\xi.
				\end{equation}
				This can be split into two parts:\\
				
				\textit{Gradient term:}
				
				\begin{equation}\label{SA11}
					\int \phi\left[-(a\psi')'\right] d\xi = -\Big[\phi,a\psi'\Big]_{\partial} + \int \phi' a \psi' d\xi.
				\end{equation}
				Boundary term vanishes under the admissible BCs, so the gradient term equals
				\begin{equation}\label{SA12}
					\int a \phi' \psi' d\xi.
				\end{equation}
				\textit{Potential term:}
				\begin{equation}\label{SA13}
					\int \phi V \psi d\xi
				\end{equation}
				is symmetric because $ V(\xi) $ is real. Therefore
				\begin{equation}\label{SA14}
					\langle \phi,\mathcal{H}\psi\rangle = \int a \phi' \psi' d\xi + \int V \phi \psi d\xi.
				\end{equation}
				Now compute $ \langle H\phi, \psi\rangle $. By the same steps we get
				\begin{equation}\label{SA15}
					\langle \mathcal{H}\phi, \psi\rangle = \int a \phi' \psi' d\xi + \int V \phi \psi d\xi.
				\end{equation}
				Comparing (\ref{SA14}) and (\ref{SA15}),
				
				\begin{equation}\label{SA16}
					\langle \phi, H\psi\rangle = \langle H\phi,\psi\rangle. 
				\end{equation}
				
				That is exactly self-adjointness. Thus, Hessian is self-adjoint in the standard $ L^2 $ inner product provided (i) $ a(\xi) $ and $ V(\xi) $ are real and sufficiently smooth, and (ii) boundary terms vanish. 
				
				\subsubsection*{Connection with the Full Hessian}
				
				If we expand (\ref{SA8}) we recover the paper’s expression:
				\begin{equation}
					-\frac{d}{d\xi}\left(a(\xi)\frac{d}{d\xi}\right) = -a(\xi)\frac{d^2}{d\xi^2} - a'(\xi)\frac{d}{d\xi}.
				\end{equation}
				With
				\begin{equation}
					a(\xi) = \mu + \Sigma(m) + \frac{3}{2}(\lambda+2\mu)(f')^2,
				\end{equation}
				this becomes precisely the ``derivative of $ ((f')^2)$ times derivative operator'' term:
				\begin{eqnarray}
					\mathcal{H}& = &-\left( \mu + \Sigma(m)\right) \frac{d^2}{d\xi^2} - \frac{3}{2}(\lambda+2\mu)\frac{d}{d\xi}\left((f')^2\frac{d}{d\xi}\right)\nonumber\\ &-& \C A + 3\C B f^2,
				\end{eqnarray}
				which is the full expression for Hessian derived in the main text.
				
				\section{Writing a uniform ``modified tanh'' equation in the whole range}
				\label{phixi}
				
				We saw that we cannot have a single exact $\tanh( \xi)$ in the brittle case. This is because the core scale is $L_c$ where as the tail scale is $\ell_0$.  However, we can write a tanh-form solution in a mathematically correct way if we consider,
				
				\begin{equation}\label{LS21}
					f(\xi) = f_0 \, \tanh\big( \phi(\xi) \big),
				\end{equation}
				
				where $\phi(\xi)$ is not assumed linear. We determine $\phi(\xi)$ from the exact first integral.\\
				
				Differentiating the above equation, we have:
				
				\begin{equation}\label{LS22}
					f' = f_0 \, \sech^2(\phi) \, \phi'.
				\end{equation}
				
				Now insert it into the exact first integral (\ref{FI13}) written in $U = \tanh \phi$, and noting that
				
				\begin{equation}\label{LS23}
					1 - U^2 = 1 - \tanh^2 \phi = \sech^2 \phi,
				\end{equation}
				then (\ref{FI13}) becomes an equation for $\phi'$:
				\begin{equation}\label{LS24}
					\frac{\epsilon_1}{2} \Big( f_0^2 \sech^4 \phi \, \phi'^2 \Big) + \frac{\epsilon_2}{4} \Big( f_0^4 \sech^8 \phi \, \phi'^4 \Big) = \frac{\C B}{4} \Big( f_0^4 \sech^4 \phi \Big).
				\end{equation}
				Divide by $\frac{f_0^4}{4} \sech^4 \phi$:
				\begin{equation}\label{LS25}
					2\epsilon_1 \frac{1}{f_0^2} \phi'^2 + \epsilon_2 \sech^4 \phi \, \phi'^4 = \C B.
				\end{equation}
				
				Now we use $f_0^2 = \C A/\C B$ to simplify $2\epsilon_1/f_0^2 = 2\sigma \C B/\C A$. Let
				
				\begin{equation}\label{LS26}
					c = \frac{2\epsilon_1\C  B}{\C A}.
				\end{equation}
				Then the exact $\phi$ equation is
				\begin{equation}\label{LS27}
					c \, \phi'^2 + \epsilon_2 \, \sech^4 \phi \, \phi'^4 = \C B.  
				\end{equation}
				This is a quadratic equation in $\phi'^2$, so we can solve it explicitly:
				Let $z = \phi'^2$. Then
				\begin{equation}\label{LS28}
					\epsilon_2\sech^4 \phi \, z^2 + c z -\C  B = 0,
				\end{equation}
				
				so
				
				\begin{equation}\label{LS29}
					\phi'^2(\phi) = \frac{-c + \sqrt{c^2 + 4\epsilon_2\C  B \, \sech^4 \phi}}{2\epsilon_2 \, \sech^4 \phi}.
				\end{equation}
				
				This gives us the general strong-slope kink in a tanh form:
				
				\begin{equation}\label{LS30}
					f(\xi) = f_0 \tanh \phi(\xi),
					\qquad
					\frac{d\phi}{d\xi} = \sqrt{ \frac{-c + \sqrt{c^2 + 4\epsilon_2 \C B \, \sech^4 \phi}}{2\epsilon_2 \, \sech^4 \phi} .
					}
				\end{equation}
				
				And then $\xi$ is obtained by quadrature:
				
				\begin{equation}\label{LS31}
					\xi - \xi_0 = \int_{0}^{\phi(\xi)} \frac{d\psi}{ \sqrt{ \frac{-c + \sqrt{c^2 + 4\epsilon_2\C B \, \sech^4 \psi}}{2\epsilon_2 \, \sech^4 \psi} } }.
				\end{equation}
				
				Note that in the tails, $\sech^4 \phi$ is very small, and $\phi' \to 1 / \ell_0$, recovering outer tanh with the length $\ell_0$. In the core of the shear-band, $\sech^4 \phi \sim 1$, and $\phi'$ scales like $1/L_c$, recovering the strong-slope core scaling. So this automatically takes care of ``modified tanh''.\\
				
				Thus, when $\alpha$ is large, the dominant balance in the core (where $\sech^4 \phi = O(1)$) gives
				
				\begin{equation}\label{LS32}
					\phi'^2 \sim \frac{\sqrt{4\alpha\C  B} - c}{2\epsilon_2} \sim \sqrt{\frac{\C B}{\epsilon_2}} \quad \Rightarrow \quad \phi' \sim \frac{1}{L_c},
				\end{equation}
				
				while in the tail $\sech^4 \phi \to 0$ gives
				
				\begin{equation}\label{LS33}
					c \, \phi'^2 \sim\C  B \quad \Rightarrow \quad \phi' \sim \sqrt{\frac{\C B}{c}} = \sqrt{\frac{\C A}{2\epsilon_1}} = \frac{1}{\ell_0}.
				\end{equation}
				
				Therefore:
				
				\begin{equation}\label{LS34}
					\phi'(\xi) \sim 
					\begin{cases}
						1 / L_c, & \text{core}, \\[4pt]
						1 / \ell_0, & \text{tails}.
					\end{cases}
				\end{equation}
				
				So the solution is tanh-like with a coordinate that behaves differently in the core and in the tails.

\end{document}